\documentclass[11pt]{article}
\usepackage[hypertexnames=false]{hyperref}
\usepackage[margin={2cm,2cm}]{geometry} 

\usepackage{sectsty}
\usepackage{abstract}
\usepackage{color}
\usepackage{graphicx}
\usepackage{multicol}
\usepackage{amsmath,amssymb,amstext,amsfonts}
\usepackage[usenames,svgnames]{xcolor}
\usepackage{float}
\usepackage{subfig}
\usepackage{caption}
\usepackage{framed}
\usepackage{placeins}
\usepackage{authblk}
\usepackage{titlesec}
\usepackage{setspace}
\usepackage{times,fancyhdr}
\usepackage[normalem]{ulem} 
\usepackage{enumerate} 
\usepackage{mathrsfs} 
\usepackage{bigints}
\usepackage{bm}
\input{undertilde} 
\usepackage{mathtools, cuted}
\usepackage{enumitem} 
\usepackage{bbm} 

\titlespacing*{\section}{0pt}{3ex plus 2ex}{1ex} 
\sectionfont{\fontsize{11}{11}\selectfont} 
\subsectionfont{\fontsize{10}{10}\selectfont} 

\hypersetup{
    colorlinks=true,
    urlcolor=SteelBlue,
    linkcolor=red,
    citecolor=blue,
}

\renewcommand{\vec}[1]{\mathbf{#1}}

\newcommand*{\Scale}[2][4]{\scalebox{#1}{$#2$}} 
\newcommand{\romansubs}{\renewcommand{\theequation}{\theparentequation \roman{equation}}} 
\newcommand{\tinymath}[1]{\mbox{\tiny{$#1$}}}
\newcommand{\elecc}{\mathcal{E}_{\mbox{\tiny{III}}}}
\newcommand{\elecb}{\mathcal{E}_{\mbox{\tiny{II}}}}
\newcommand{\eleca}{\mathcal{E}_{\mbox{\tiny{I}}}}
\newcommand{\potc}{\mathcal{W}_{\mbox{\tiny{III}}}}
\newcommand{\potb}{\mathcal{W}_{\mbox{\tiny{II}}}}
\newcommand{\pota}{\mathcal{W}_{\mbox{\tiny{I}}}}
\newcommand{\pc}{\mathsf{E}_{{\mbox{\tiny{III}}}}}
\newcommand{\pb}{\mathsf{E}_{{\mbox{\tiny{II}}}}}
\newcommand{\pa}{\mathsf{E}_{{\mbox{\tiny{I}}}}}
\newcommand{\cee}{\mathsf{A}_{{\mbox{\tiny{III}}}}}
\newcommand{\bee}{\mathsf{A}_{{\mbox{\tiny{II}}}}}
\newcommand{\ay}{\mathsf{A}_{{\mbox{\tiny{I}}}}}
\newcommand{\eleccdot}{\dot{\mathcal{E}}_{\mbox{\tiny{III}}}}
\newcommand{\elecbdot}{\dot{\mathcal{E}}_{\mbox{\tiny{II}}}}

\newcommand{\potcdot}{\dot{\mathcal{W}}_{\mbox{\tiny{III}}}}
\newcommand{\potbdot}{\dot{\mathcal{W}}_{\mbox{\tiny{II}}}}

\newcommand{\pcdot}{\dot{\mathsf{E}}_{{\mbox{\tiny{III}}}}}
\newcommand{\pbdot}{\dot{\mathsf{E}}_{{\mbox{\tiny{II}}}}}

\newcommand{\ceedot}{\dot{\mathsf{A}}_{{\mbox{\tiny{III}}}}}
\newcommand{\beedot}{\dot{\mathsf{A}}_{{\mbox{\tiny{II}}}}}

\newcommand{\subsX}{_{\mbox{\tiny{$X$}}}}
\newcommand{\qxx}{q_{\mbox{\tiny{$X\hspace{-0.08cm}X$}}}}
\newcommand{\qxxdot}{\dot{q}_{\mbox{\tiny{$X\hspace{-0.08cm}X$}}}}
\newcommand{\amin}{a_{\mbox{\tiny{min}}}}
\newcommand{\ymfront}{\frac{\epsilon_{abc}\epsilon^{ijk}E_{j}^{\;b}E_{k}^{\;c}}{8 \left(\mbox{det}(E)\right)^{3/2}}  \epsilon_{def}\epsilon_{i}^{\;mn}E_{m}^{\;e}E_{n}^{\;f}}
\newcommand{\tinyrmsub}[1]{\mbox{{\tiny{#1}}}}

\newcommand{\PRLsep}{\noindent\makebox[\linewidth]{\resizebox{0.750\linewidth}{1pt}{$\blacklozenge$}}\bigskip}

\begin{document}
\pagestyle{fancy}
\fancyhead{} 
\fancyhead[OR]{\thepage}
\fancyhead[OC]{{\small{\textsf{LOOP QUANTUM CORRECTED EYM BLACK HOLES}}}}
\fancyfoot{} 
\renewcommand\headrulewidth{0.5pt}
\addtolength{\headheight}{2pt} 
\global\long\def\tdud#1#2#3#4#5{#1_{#2}{}^{#3}{}_{#4}{}^{#5}}
\global\long\def\tudu#1#2#3#4#5{#1^{#2}{}_{#3}{}^{#4}{}_{#5}}

\twocolumn

\title{\vspace{-2cm}\hspace{-0.0cm}\rule{\linewidth}{0.2mm}\\ \bf{\Large{\textsf{Loop Quantum Corrected Einstein Yang-Mills Black Holes}}}}


\author[1]{\small{Mason Protter}\thanks{\href{mailto:protter@ualberta.ca}{protter@ualberta.ca}}}%

\author[2,3]{\small{Andrew DeBenedictis}\thanks{\href{mailto:adebened@sfu.ca}{adebened@sfu.ca}}}%


\affil[1]{\footnotesize{\it{Department of Physics, The University of Alberta}}\\
\footnotesize{\it{Edmonton, AB, T6G 2E1, Canada}}}
\affil[2]{\footnotesize{\it{The Pacific Institute for the Mathematical Sciences}} \protect\\
\footnotesize{and}}
\affil[3]{\footnotesize{\it{Department of Physics, Simon Fraser University}}\\
\footnotesize{\it{8888 University Drive, Burnaby, BC, V5A 1S6, Canada}}}

\date{\vspace{-0.8cm}(\footnotesize{April 8, 2018})} 
\twocolumn[ 
  \begin{@twocolumnfalse}  
  \begin{changemargin}{1.75cm}{1.75cm} 
\maketitle
\end{changemargin}
\vspace{-1.0cm}
\begin{changemargin}{1.5cm}{1.5cm} 
\begin{abstract}
{\noindent\small{In this brief paper we study the homogeneous interiors of black holes possessing SU(2) Yang-Mills fields subject to corrections inspired by loop quantum gravity. The systems studied possess both magnetic and induced electric Yang-Mills fields. We consider the system of equations both with and without Wilson loop corrections to the Yang-Mills potential. The structure of the Yang-Mills Hamiltonian along with the restriction to homogeneity allows for an anomaly free effective quantization. In particular we study the bounce which replaces the classical singularity and the behavior of the Yang-Mills fields in the quantum corrected interior, which possesses topology $R\times S^{2}$. Beyond the bounce the magnitude of the Yang-Mills electric field asymptotically grows monotonically. This results in an ever expanding $R$ sector even though the two-sphere volume is asymptotically constant. The results are similar with and without Wilson loop corrections on the Yang-Mills potential.}}
\end{abstract}
\noindent{\footnotesize PACS(2010): 04.70.Dy\;\; 04.60.Pp\;\; 12.15.-y}
\rule{\linewidth}{0.2mm}
\end{changemargin}
\end{@twocolumnfalse} 
]
\saythanks 
\vspace{0.5cm}
{\setstretch{0.9} 
\section{Introduction}
It may be argued that black holes are one of the most fascinating objects predicted by gravitational field theory. Black holes are likely not just theoretical musings but are believed to be abundant in our universe \cite{ref:astrobh1}-\cite{ref:astrobh3}, with the most recent compelling evidence coming from the fascinating new field of gravitational wave astronomy (see \cite{ref:gravwave} for an overview). The indirect evidences of black holes in our universe all seem to vindicate the predictions of general relativity, and it is difficult to argue that general relativity is not our best theory of gravitation to date. As successful and aesthetically pleasing as general relativity is, it does present some issues when making predictions about black holes. One issue that has garnered much attention in the literature at least since Oppenheimer and Snyder's seminal work \cite{ref:oppsnyd} is that of the singularity inside the black hole. Although perhaps a troubling prediction, it is often thought that at such high energies quantum gravity effects would begin to dominate and that a quantum theory of gravity would alleviate the singularity problem. However, general relativity and related theories have to date eluded a fully satisfactory quantization. There are many reasons for this but one central issue is that the diffeomorphism invariance of the theory, in the form of background independence, makes such theories difficult to quantize by standard techniques \cite{ref:isham}, \cite{ref:smolin}. There are several candidate quantum gravity theories which attempt to address this issue (see \cite{ref:diffeocausalset}, \cite{ref:ambjornCDT}, \cite{ref:backindLQG} and references therein) one of which is the theory of Loop Quantum Gravity \cite{ref:lehtinen}. In the loop approach the singularity issue, both in a cosmological setting and in black holes, has been dealt with on many levels {\cite{ref:modestobhsing2}-\cite{ref:LQCsingreview}. Interestingly, there is an implication of signature change in some models due to variable deformation \cite{ref:brahma1}, \cite{ref:brahma2}. Due to the difficulty of the problem in general, one often appeals to effective techniques. One such technique gives rise to a holonomy corrected effective theory. Here one replaces the connection variable of loop quantum gravity, $A^{i}_{\;a}$, with the holonomy of the connection inspired by the non-canonical algebra of the quantized theory \cite{ref:gambinialgebra}. At low-order this replacement usually takes the form (see appendix)
\begin{equation}
 A^{i}_{\;a} \rightarrow \frac{\sin\left(A^{i}_{\;a} \delta\right)}{\delta}\;, \label{eq:holreplacement}
\end{equation}
where $\delta$ is related to the length of the holonomy path (more details are provided later). We adopt the convention that indices $a, b, c, $ etc. are spatial indices whereas indices $i, j, k,...$ and $I, J, K,...$ are $su(2)$ indices in the gravitational sector and Yang-Mills sector respectively. We study the system both with and without finite size Wilson loop corrections to the Yang-Mills field.

One issue that can arise with such deformations is that they can introduce an anomaly in the algebra of constraints. There are two important issues regarding the constraint algebra of diffeomorphism invariant theories \cite{ref:bojowaldalgebra}, \cite{ref:achouralgebra}; The algebra must remain first-class, and the algebra must reduce to the usual general relativity algebra in the limit that the variable deformations go to zero. These issues have been addressed in a number of interesting and sophisticated ways \cite{ref:bojowaldalgebra}-\cite{ref:achouralgebra2}. More general deformations of gravitational variables respecting an acceptable constraint algebra can be found in \cite{ref:tomasetal}. One scenario where the anomaly issue does not arise with holonomy corrections is in purely homogeneous scenarios \cite{ref:achouralgebra}. Essentially, the automatic satisfaction of the smeared vector constraint, along with the trivialization of the integrals in the algebra due to homogeneity leads to a trivial algebra. 

This is why we limit ourselves to homogeneous black hole interiors, sometimes known in the literature as ``T-spheres''. T-spheres are of interest not only because the Schwarzschild black hole's interior is a T-sphere, but they have also been studied under more generic conditions \cite{ref:ruban1}- \cite{ref:zaslatsph}. They have also been used to study the singularity issue in other theories of gravity \cite{ref:aftergood}. The T-spheres make up a generalization of the original Kantowski-Sachs metric \cite{ref:kantsachs}. In the realm of loop quantum gravity, the Schwarzschild interior has been studied in a number of interesting works \cite{ref:AandB}, \cite{ref:modestobhsing} - \cite{ref:modestoint} and there have been extensions to homogeneous interiors with cosmological constant and non-trivial topologies \cite{ref:BKD}. The Reissner-Nordstr\"{o}m scenario has been studied in \cite{ref:taslimitehrani}, \cite{ref:tibrewalaeinstmax}, \cite{ref:tehrani}. Recently, the singularity avoidance issue in Schwarzschild spacetime has been revisited in \cite{ref:schwrevisit}, pointing out some issues, and some interesting general results with matter have been attained \cite{ref:singh2}.

We adopt the interior line element of the following form \footnote{In this coordinate chart the Schwarzschild line element has the form $\mathrm{d}s^{2}=-\frac{\mathrm{d}T^{2}}{\frac{2M}{T}-1} + \left(\frac{2M}{T}-1\right)\mathrm{d}X^{2}+T^{2}\left(\mathrm{d}\theta^{2} + \sin^{2}\theta\, \mathrm{d}\phi^{2}\right)$ with the coordinate restriction $0 < T < 2M$.}
\begin{eqnarray}
\mathrm{d}s^{2}=&-N^{2}(T)\,\mathrm{d}T^{2}+\qxx(T)\,\mathrm{d}X^{2} \nonumber \\
&+T^{2}\left(\mathrm{d}\theta^{2} + \sin^{2}\theta\, \mathrm{d}\phi^{2}\right)\,, \label{eq:intline}
\end{eqnarray}
and also introduce a homogeneous Yang-Mills field in the black hole spacetime.

The arena of Einstein Yang-Mills (EYM) theory is extremely rich, due to the interplay between the complexity of both general relativity and Yang-Mills theories.
Non black hole globally regular solutions exits which are asymptotically flat \cite{ref:BM}-\cite{ref:wasserman}. These are magnetic and are characterized by the number of zero nodes of the Yang-Mills potential. Interestingly, a similar behavior had been noted previously in the Einstein-Maxwell-Klein-Gordon system by Das and Coffman \cite{ref:dascoff}. Aside from the regular solutions black hole solutions also exist \cite{ref:volgal}-\cite{ref:donets}. These have been characterized as S-type (Schwarzschild) and RN-type (Reissner-Nordstr\"{o}m), which describes their asymptotic structure near the singularity \cite{ref:donets}, \cite{ref:volkovreview}. There also exist EYM black hole solutions which possess highly oscillating metric functions in the interior region \cite{ref:donets}, \cite{ref:volkovreview}. The study of solutions has been extended to include the presence of a Higgs fields \cite{ref:galthiggs}-\cite{ref:jiahiggs}, and those with non-asymptotically flat exteriors \cite{ref:winstanley1}-\cite{ref:breitlambda}. Symmetries have been also relaxed from spherical symmetry to axial symmetry \cite{ref:pertaxial}, \cite{ref:kleiaxial}. Other interesting issues such as stability, nonminimal coupling, scaling, dyons and monopoles have also been investigated (see \cite{ref:bizondyon}-\cite{ref:criticalmaliborsk} and references therein).  Detailed reviews of EYM theory may be found in \cite{ref:volkovreview}, \cite{ref:samarkandrev}, \cite{ref:richness}.

Our study here is limited to comparing purely classical evolutions inside the black hole to quantum corrected evolutions. The node structure of the Yang-Mills potentials is not studied (and in fact cannot be since some of the nodes presumably occur in the exterior region, where our analysis cannot be extended). We also do not study the stability properties of such black holes. The question we wish to address is the one of what the effects of loop quantum corrections are on both the spacetime geometry and the Yang-Mills fields and a comparison to purely classical theory.


\section{EYM black hole interiors in Ashtekar variables}\label{sec:EYMBH}
In the variables appropriate for the real connection version of loop quantum gravity one has a densitized triad, $E_{i}^{\;a}$, and $su(2)$ connection, $A^{i}_{\;a}$, as the canonical gravitation variables. The classical Poisson bracket is given by
\begin{equation}
 \left\{A^{i}_{\;a}(x),\,E_{j}^{\;b}(y)\right\}=\kappa \gamma \delta^{i}_{\;j} \delta_{a}^{\;b}\delta(x,\,y)\,. \label{eq:aebracket}
\end{equation}
Here $\kappa=8\pi$ and $\gamma$ is a parameter known as the Immirzi parameter whose value is usually set via black hole entropy calculations \cite{ref:entrev1}-\cite{ref:highergenusent}. These variables are related to the more familiar ADM variables via:
\begin{subequations}
\romansubs
{\allowdisplaybreaks\begin{align}
A^{i}_{\;a}&=\Gamma^{i}_{\;a} + \gamma K^{i}_{\;a}\,, \label{eq:AKreln}\\
E_{i}^{\;a}E_{j}^{\;b}\delta^{ij}&=q\,q^{ab}\,, \label{eq:tetmetreln}
\end{align}}
\end{subequations}
with $\Gamma^{i}_{\;a}$ the spin-connection:
\begin{equation}
 \Gamma^{i}_{\;a}:=2\epsilon^{ij}_{\;\;\;k} e_{j}^{\;b}\left[\partial_{[a} e^{k}_{\;b]}+\frac{1}{2} \delta^{kl}\delta_{mn}e^{c}_{\;l}e^{m}_{\;a}\partial_{b}e^{n}_{\;c}\right]\,, \label{eq:Gamma}
 \end{equation}
$q^{ab}$ the (inverse) 3-metric, and $q$ the metric determinant. The quantity $K^{i}_{\;a}$ is the densitized extrinsic curvature,
\begin{equation}
 K^{i}_{\;a}:=\frac{E_{j}^{\;b}\delta^{ij}K_{ab} }{\sqrt{\mbox{det}(E)}}\,, \label{eq:densext}
\end{equation}
$K_{ab}$ being the usual extrinsic curvature and $E$ the determinant of $E_{i}^{\;a}$. In (\ref{eq:Gamma}) the regular tetrad, $e_{j}^{\;a}$, is generally to be rewritten in terms of the densitized triad:
\begin{subequations}
\romansubs
{\allowdisplaybreaks\begin{align}
e^{i}_{\;a}&= \frac{1}{2\sqrt{\mbox{det}(E)}}\epsilon^{ijk} \epsilon_{abc} E_{j}^{\;b}E_{k}^{\;c}\,, \label{eq:tettetreln1}\\ 
e_{i}^{\;a}&= \frac{E_{i}^{\;a}}{\sqrt{\mbox{det}(E)}}\,. \label{eq:tettetreln2}
\end{align}}
\end{subequations}

In our study, we reduce the system to spherical symmetry and hence adopt the following spherically symmetric gravitational densitized triad and connection pair:
\begin{subequations}
\romansubs
{\allowdisplaybreaks\begin{align}
E_{i}^{\;a}\tau^{i}\partial_{a}=&\pc\tau^{3}\partial\subsX+ \left(\pa\tau^{1}+\pb\tau^{2}\right)\sin\theta\,\partial_{\theta} \nonumber \\
& \quad \; \qquad +\left(\pa\tau^{2}-\pb\tau^{1}\right) \partial_{\phi}\,, \label{eq:ge_ansatz} \\[0.1cm]
A^{i}_{\;a}\tau_{i}dx^{a}=&\Scale[0.95]{\cee(T)\tau_{3}\mathrm{d}X+\left(\ay\tau_{1}+\bee\tau_{2}\right)\mathrm{d}\theta} \nonumber \\
&\Scale[0.95]{+\left(\ay\tau_{2}-\bee\tau_{1}+\cot\theta\,\tau_{3}\right)\sin\theta\,\mathrm{d}\phi\,,} \label{eq:ga_ansatz}
\end{align}}
\end{subequations}
where the $\tau^{i}$ are the (spherical) $SU(2)$ generators.

A similar ansatz holds for the $SU(2)$ Yang-Mills electric field and potentials:
\begin{subequations}
\romansubs
{\allowdisplaybreaks\begin{align}
\mathscr{E}_{I}^{\;a}\tau^{I}\partial_{a}=& \elecc\tau^{3}\partial\subsX +\left(\eleca\tau^{1}+\elecb\tau^{2}\right)\sin\theta\,\partial_{\theta} \nonumber \\
& \quad \; \qquad +\left(\eleca\tau^{2}-\elecb\tau^{1}\right) \partial_{\phi}\,, \label{eq:elec_ansatz} \\[0.1cm]
W^{I}_{\;a}\tau_{I}dx^{a}=&\Scale[0.90]{\potc\tau_{3}\mathrm{d}X+\left(\pota\tau_{1}+\potb\tau_{2}\right)\mathrm{d}\theta} \nonumber \\
&\Scale[0.90]{+\left(\pota\tau_{2}-\potb\tau_{1}+\cot\theta\,\tau_{3}\right)\sin\theta\,\mathrm{d}\phi\,.} \label{eq:pot_ansatz}
\end{align}}
\end{subequations}
In the above, all functions ($\mathsf{E}$, $\mathsf{A}$, $\mathcal{E}$, $\mathcal{W}$) are functions of the interior time, $T$, only. As discussed in the introduction, the above momentum-configuration conjugate pairs do not produce an anomaly in the resulting algebra of gravitational constraints.

From the set of variables defined in (\ref{eq:ge_ansatz}-\ref{eq:pot_ansatz}) one may form the EYM action in canonical form: \vspace{-0.2cm}
\begin{align}
I=&I_{\mbox{\tiny{grav}}}+I_{\mbox{\tiny{YM}}}  \nonumber \\
=&\Scale[0.90]{ \bigintss_{\mathbb{R}} dt \bigintss_{\Sigma} d^{3}x \bigg\{\left[\frac{1}{\kappa}  E_{i}^{\;a} \dot{A}^{i}_{\;a} + \frac{1}{\tiny{Q}^{2}}\mathscr{E}_{I}^{\;a}\dot{W}^{\tinymath{I}}_{\;a} - \mathcal{H} \right] \bigg\}} \label{eq:totact} \\\nonumber \\
=&\Scale[0.90]{\bigintss_{\mathbb{R}} dt \bigintss_{\Sigma} d^{3}x \bigg\{\frac{1}{\kappa} \left[ E_{i}^{\;a} \dot{A}^{i}_{\;a}  -N^{b} V_{b}  - N S_{\mbox{\tiny{grav}}} -\lambda^{i} G_{i}\right]} \nonumber \\
&\Scale[0.95]{+\frac{1}{Q^{2}}\left[\mathscr{E}_{I}^{\;a}\dot{W}^{\tinymath{I}}_{\;a} +W^{\tinymath{I}}_{\;T}D_{a}\mathscr{E}_{\tinymath{I}}^{\;a} -N^{a}\mathscr{E}_{\tinymath{I}}^{\;b} F^{\tinymath{I}}_{\;ab}+ N S_{\mbox{\tiny{YM}}} \right]\bigg\} }\,,  \nonumber
\end{align}
with $Q$ the Yang-Mills coupling, which we will set to $1$ for convenience. Here $N^{a}$ is the ADM shift vector and $N$ the lapse function, and $\lambda^{i}$ an $su(2)$ valued Lagrange multiplier, related to the time component of the gravitational connection. The other quantities appearing are defined as follows:
\begin{subequations}
\romansubs
{\allowdisplaybreaks\begin{align}
&G_{i}:=\partial_{a} E_{i}^{\;a} +\epsilon_{ij}^{\;\;\;k}A^{j}_{\;a}E_{k}^{\;a}\, , \label{eq:gaussconst} \\
&V_{b}:=E_{i}^{\;a}F^{i}_{\;ab}-(1+\gamma^{2})K^{i}_{\;b}G_{i}\,, \label{eq:vecconst} \\
&\Scale[0.93]{S_{\mbox{\tiny{grav}}}:=\frac{E_{i}^{\;a}E_{j}^{\;b}}{\sqrt{\mbox{det}(E)}} \left[\epsilon^{ij}_{\;\;\;k}F^{k}_{\;ab}-2(1+\gamma^{2})K^{i}_{\;[a}K^{j}_{\;b]}\right]} \,,\label{eq:scalconst} \\
&D_{a}\mathscr{E}_{\tinymath{I}}^{\;a}:=\partial_{a} \mathscr{E}_{\tinymath{I}}^{\;a} +\epsilon_{\tinymath{IJ}}^{\;\;\;\tinymath{K}}W^{\tinymath{J}}_{\;a}\mathscr{E}_{\tinymath{K}}^{\;a}\,,\label{ymvecconst}\\
&S_{\mbox{\tiny{YM}}}:=\Scale[0.95]{\ymfront} \nonumber \\
&\quad\qquad\times \left[\mathscr{E}_{\tinymath{I}}^{\;a} \mathscr{E}_{\tinymath{J}}^{\;d} + \mathscr{B}_{\tinymath{I}}^{\;a} \mathscr{B}_{\tinymath{J}}^{\;d}\right] \delta^{\tinymath{IJ}}\,,  \label{eq:ymscalconst}
\end{align}}
\end{subequations}
and $F^{i}_{\;ab}:=\partial_{a}A^{i}_{\;b}-\partial_{b}A^{i}_{\;a}+\epsilon^{i}_{\;jk}A^{j}_{\;a}A^{k}_{\;b}$ with a similar definition for the curvature of the Yang-Mills potential, $F^{\tinymath{I}}_{\;ab}$. Quantities linear in the shift vector, $N^{a}$, are often called the vector constraint, and quantities linear in the lapse make up the scalar constraint (sometimes referred to as the Hamiltonian constraint). The $\mathscr{B}_{\tinymath{I}}^{\;a}$ are magnetic field components, which must be written in terms of the canonical configuration variable, $W^{\tinymath{I}}_{\;a}$:
\begin{equation}
 \mathscr{B}_{\tinymath{I}}^{\;a}:=\frac{1}{2}\epsilon^{abc}F^{\tinymath{J}}_{\;bc}\delta_{\tinymath{IJ}}\,. \label{eq:mag}
\end{equation}

Using the ansatz (\ref{eq:ge_ansatz}-\ref{eq:pot_ansatz}) in (\ref{eq:gaussconst} - \ref{eq:ymscalconst}) yields the following for the Gauss constraints:
\begin{subequations}
\romansubs
{\allowdisplaybreaks\begin{align}
G_{i}=&8\pi\left[\ay\pb-\bee\pa\right]\delta_{i3}=0\,, \label{eq:spheregravgauss} \\
D_{a}\mathscr{E}_{\tinymath{I}}^{\;a}=&8\pi\left[\pota\elecb-\potb\eleca\right]\delta_{I3}=0\,. \label{eq:sphereymgauss}
\end{align}}
\end{subequations}
These constraints enforce the fact that in the $\mbox{I}-\mbox{II}$ subspace it is only the magnitude of a ``vector'' (eg. $\sqrt{\pa^{2}+\pb^{2}}$, etc.) which is relevant for the equations of motion. Any rotation of these vectors is equally acceptable.

In polymer methods, the constraints (\ref{eq:spheregravgauss},ii) are fixed before effective quantization. This is due to the gauge conditions they employ. (Also \cite{ref:achouralgebra}, \cite{ref:achouralgebra2} and references therein). We therefore satisfy these constraints with the choice
\begin{equation}
 \pa=0=\ay \;\;\;\mbox{and}\;\;\; \eleca=0=\pota\,. \label{eq:gausssatisfaction}
\end{equation}
With this choice, and (\ref{eq:tetmetreln}), a small calculation reveals that
\begin{equation}
 q_{\theta\theta}=\pc\;\;\mbox{and}\;\;\ \qxx=\frac{\pb^{2}}{\pc}\,, \label{eq:explicittetmetreln}
 \end{equation}
so that the line-element (\ref{eq:intline}) may be written as:
\begin{equation}
\Scale[0.89]{\mathrm{d}s^{2}=-N^{2}\,\mathrm{d}T^{2}+\frac{\pb^{2}}{\pc}\,\mathrm{d}X^{2}+\pc\left(\mathrm{d}\theta^{2} + \sin^{2}\theta\, \mathrm{d}\phi^{2}\right).} \label{eq:tetradline}
\end{equation}

With gauge fixing and the automatic satisfaction of the (smeared) vector constraint due to homogeneity in (\ref{eq:ge_ansatz})-(\ref{eq:pot_ansatz}), we can now write down the Hamiltonian, $H=\int d^{3}x\,\mathcal{H}$, required for the evolution (setting the (arbitrary) upper-limit of the $X$ integral to $1$):
\begin{align}
 H=&\frac{N}{2\pi\gamma^{2}\sqrt{\pc}}\left[(\bee^2+\gamma^2) \pb +2\bee\cee\pc\right] \nonumber \\
 &+\frac{N}{4\gamma^{2}\pb\pc^{3/2}} \left[\pb^{2}\left(\potb^{2}-1\right)^{2} \right. \nonumber \\
 &\Scale[0.99]{\left. +2\potb^{2}\potc^{2}\pc^{2}+2\elecb^{2}\pc^{2}+\elecc^{2}\pb^{2}\right]}. \label{eq:htot}
\end{align}
(The dimensions of the gravitational and Yang-Mills contribution are equivalent due to having set $G=1=c$).

In the literature on spherically symmetric Einstein Yang-Mills, the metric (\ref{eq:intline}) is often taken to have the form \cite{ref:volgal}, \cite{ref:donets}, \cite{ref:improved}
\begin{equation}
\Scale[0.93]{ \mathrm{d}s^{2}= -\frac{T^{2}}{|\Delta|} \mathrm{d}T^{2} + \frac{|\Delta|}{T^{2}}\sigma^{2}\,\mathrm{d}X^{2}+T^{2}\left(\mathrm{d}\theta^{2} + \sin^{2}\theta\, \mathrm{d}\phi^{2}\right)\,,} \label{eq:ymline}
\end{equation}
where we have re-written the traditional form in the interior coordinate chart. The functions $\Delta$ and $\sigma$ solely depend on $T$ and the T-domain corresponds to the inequality $\Delta < 0$.

Comparing (\ref{eq:ymline}) with (\ref{eq:tetradline}) one arrives at the following relationship:
\begin{equation}
 N=\frac{\sqrt{\pc}}{\pb}\sigma\,, \label{eq:Noftet}
 \end{equation}
where in (\ref{eq:Noftet}) the $T$ dependence in $\sigma$ is to be replaced with $T\rightarrow\sqrt{\pc}$ (as dictated by the first equation in (\ref{eq:explicittetmetreln})). Strictly speaking, the form on $N$ is not of great importance to the evolution as it plays the role of a Lagrange multiplier in the action.

The equations of motion are
\begin{subequations}
\romansubs
{\allowdisplaybreaks\begin{align}
{\ceedot}=&2\frac{\partial H}{\partial \pc}\,, \label{eq:ceqn} \\
{\pcdot}=&-2\frac{\partial H}{\partial \cee}\,, \label{eq:pceqn} \\
{\beedot}=&\frac{\partial H}{\partial \pb}\,, \label{eq:beqn} \\
{\pbdot}=&-\frac{\partial H}{\partial \bee}\,, \label{eq:pbeqn} \\
{\potbdot}=&\frac{1}{4\pi}\frac{\partial H}{\partial \elecb}\,, \label{eq:potbeqn} \\
{\elecbdot}=&-\frac{1}{4\pi}\frac{\partial H}{\partial \potb}\,, \label{eq:elecbeqn} \\
{\potcdot}=&\frac{1}{4\pi}\frac{\partial H}{\partial \elecc}\,, \label{eq:potceqn} \\
{\eleccdot}=&-\frac{1}{4\pi}\frac{\partial H}{\partial \potc}\,. \label{eq:elecceqn}
\end{align}}
\end{subequations}
The explicit equations are rather lengthy, so are omitted, but they can easily be obtained by performing the differentiations above. At this stage, save for requiring initial conditions, the system of equations can be solved.

\subsection{Classical vs holonomy corrected evolution}\label{subsec:evolve}
In this subsection we evolve the system of equations (\ref{eq:ceqn}-viii) in both the classical case and the holonomy corrected one. In both scenarios we start the evolution away from the classical singularity where we expect that the classical general relativity solutions to be valid and therefore use the same (purely classical) initial conditions for both the classical and holonomy corrected evolutions. Due to the complexity of the Einstein Yang-Mills system there is a dearth of analytic solutions available in the literature. However, there are a number of very interesting computational studies of the EYM equations from which one may determine initial conditions, and we therefore utilize a similar initial ansatz as in \cite{ref:donets}. For the Yang-Mills potentials the following is chosen:
\begin{equation}
 \potb\neq 0, \;\; \potc=0\,. \label{eq:specificpotansatz}
\end{equation}
It is interesting to note that although (\ref{eq:specificpotansatz}) corresponds to a magnetic ansatz in the static case, there exists an induced electric field in time dependent domains such as studied here via
\begin{equation}
 \mathscr{E}_{\tinymath{L}}^{\;a}=\frac{E_{i}^{\;a}E_{j}^{\;b}\delta^{ij}}{\sqrt{\mbox{det}(E)}}\hat{n}^{\mu}F^{\tinymath{K}}_{\;\;\mu b}\delta_{\tinymath{KL}}\,, \label{eq:inducedelec}
\end{equation}
with $\hat{n}^{\mu}$ the unit normal vector to the $\Sigma$ hypersurface. In this case the induced electric field given by (\ref{eq:inducedelec}) is calculated as
\begin{subequations}
\romansubs
{\allowdisplaybreaks\begin{align}
\elecb&=\frac{\sqrt{|\Delta|}\sigma}{N T} \potbdot \,, \label{eq:specificelecb}\\
\elecc&=0  \,. \label{eq:specificelecc}
\end{align}}
\end{subequations}
These are used to calculate the initial electric field values. We also need to know the gravitational connection components, $\bee$ and $\cee$, in order to determine their initial values. These can be gotten from (\ref{eq:AKreln}) and (\ref{eq:Gamma} - \ref{eq:tettetreln2}) as
\begin{subequations}
\romansubs
{\allowdisplaybreaks\begin{align}
\bee&=\pm\frac{\gamma}{N} \,, \label{eq:specificb} \\
\cee&=\pm \frac{\gamma}{2N}\frac{\qxxdot}{\sqrt{\qxx}}\,. \label{eq:specificc}
\end{align}}
\end{subequations}

In order to evolve the system using the above relationships, the following data at the initial time is needed: $\sigma_{\mbox{\tiny{init}}}$, $\dot{\sigma}_{\mbox{\tiny{init}}}$, $\Delta_{\mbox{\tiny{init}}}$, $\dot{\Delta}_{\mbox{\tiny{init}}}$, ${\mathcal{W}}_{\tiny{\mbox{II\,init}}}$, $\dot{\mathcal{W}}_{\mbox{\tiny{II\,init}}}$ and of course the initial time itself. All required initial quantities can be formed from these. (The requirement of the time derivative quantities may seem peculiar in the Hamiltonian formalism of a second-order theory but they are only needed so that one may calculate ${\elecb}_{\mbox{\tiny{init}}}$ and ${\cee}_{\mbox{\tiny{init}}}$ via (\ref{eq:specificelecb}) and (\ref{eq:specificc}).) As mentioned previously, these initial conditions are used for both the classical as well as quantum corrected evolution since the initial time is not near the classical singular point where quantum effects are thought to dominate.

A quantity of particular interest is $\pc$ as, from (\ref{eq:tetradline}), it is the quantity which determines the volumes of two-spheres inside the black hole. If $\pc \rightarrow 0$ then, as time progresses, the two-sphere volume shrinks to zero and hence a singularity of some type will be present. (Although if energy conditions are violated in the classical case the singularity \emph{might} not impede geodesic completeness.)

For the quantum corrected scenarios we replace the connection components, $\bee$ and $\cee$, in (\ref{eq:htot}) by their holonomy corrected counter-parts as in (\ref{eq:holreplacement}). That is, for the particular case under study:
\begin{subequations}
\romansubs
{\allowdisplaybreaks\begin{align}
\bee \rightarrow& \frac{\sin\left(\bee \delta\subsX\right)}{\delta\subsX}\,, \label{eq:holob}\\
\cee \rightarrow& \frac{\sin\left(\cee \delta_{\theta}\right)}{\delta_{\theta}}\,. \label{eq:holoc}
\end{align}}
\end{subequations}
As mentioned in the introduction the quantities $\delta\subsX$ and $\delta_{\theta}$ depend on the proper length along the coordinate directions the holonomy path is taken \cite{ref:mubarprime}. We utilize here the ``$\overline{\mu}^{\prime}$ scheme'' introduced in \cite{ref:BandV}. In early works on singularity avoidance in effective LQG $\delta$ was taken to be constant. However, it was later shown that a constant $\delta$ does not yield a good semi-classical limit \cite{ref:musemiclass}, and that relating $\delta$ to the proper length yields a correct semi-classical limit. The $\overline{\mu}^{\prime}$ specifically has been shown to generally lead to dynamics which are independent of the size of the $X$ integral in the Hamiltonian (\ref{eq:htot}) \cite{ref:chiouKS}. In the $\overline{\mu}^{\prime}$ scheme $\delta$ is computed as follows:
\begin{subequations}
\romansubs
{\allowdisplaybreaks\begin{align}
a_{\mbox{\tiny{$X$}}\theta}=&\sqrt{\qxx}\delta\subsX\, T\delta_{\theta}\,, \label{eq:aminXtheta}\\
a_{\theta\phi}=&\left(T\delta_{\theta}\right)^{2} \,, \label{eq:aminthetaphi}
\end{align}}
\end{subequations}
where $a_{bc}$ indicate a small area element in the $b-c$ plane. In this case, using (\ref{eq:explicittetmetreln}), the quantities are
\begin{equation}
 \delta\subsX=\frac{(\ell_{1})^{2}\sqrt{\pc}}{\ell_{2} \pb}\;\;\; \mbox{and} \;\;\; \delta_{\theta}=\frac{\ell_{2}}{\sqrt{\pc}}\,,
\end{equation}
with $\ell_{1}:=\sqrt{a_{\mbox{\tiny{$X$}}\theta}}$\, and\, $\ell_{2}:=\sqrt{a_{\theta\phi}}$\,. Often the choice is made to take $\ell_{1}=\ell_{2}=\sqrt{\amin}$, where $\amin$ is the minimum value of Loop Quantum Gravity's area spectrum. However, there is no definitive reason to do so. The qualitative result (bounce) remains for all non-zero choices, although the deviation from the purely classical evolution is noticeable earlier for larger values.

It is also possible to implement finite plaquette size effects on the Yang-Mills potentials from a Wilson loop approach. The rationale is that the derivation of the holonomy corrections (see appendix for a brief over\-view) is independent of whether the $SU(2)$ field-strength tensor is a gravitational one or not. Here we will consider both corrected and uncorrected Yang-Mills potentials. First we will consider \emph{no} Wilson loop corrections. The argument, inspired by lattice gauge theory, for ignoring these corrections at low-order is as follows. Consider a Taylor expansion of the the potential $\mathcal{W}_{\tinyrmsub{B}}$ (B$\in\{\mathrm{II},\,\mathrm{III}\}$) on a finite element sized lattice \cite{ref:schwartzbook}, in this case presumably due to the discrete area spectrum introduced by loop quantum gravity:
\begin{equation}
\mathcal{W}_{\tinyrmsub{B}}(n+a)=\mathcal{W}_{\tinyrmsub{B}}(n)+\delta\,\partial_{a}\mathcal{W}_{\tinyrmsub{B}}+\mathcal{O}(\delta^{2})\,, \label{eq:pothop}
\end{equation}
where $\delta$ represents the lattice spacing in the direction of the ``hop'' (in this case the ``$a$'' direction). It may have different values for different hopping directions. As a crude approximation let us take our Yang-Mills field to be defined as its averaged value between two neighboring lattice points, although the result of the argument is independent of the exact position one chooses. Then, using (\ref{eq:pothop}),
\begin{align}
 \mathcal{W}_{\tinyrmsub{B}}(\vec{x})&=\frac{\mathcal{W}_{\tinyrmsub{B}}(n)+\mathcal{W}_{\tinyrmsub{B}}(n+a)}{2} \nonumber \\
 &= \mathcal{W}_{\tinyrmsub{B}}(n) + \frac{\delta}{2} \partial_{a}\mathcal{W}_{\tinyrmsub{B}}(n) + \mathcal{O}(\delta^{2})\,.\label{eq:avgpot}
\end{align}
Since the $\mathcal{W}_{\tinyrmsub{B}}$ are homogeneous, the partial derivative correction in (\ref{eq:avgpot}) will not contribute. Also due to homogeneity, the field at the midpoint, $\vec{x}$, is the same as at the initial lattice point, $n$. The electric field similarly does not pick up a correction at this level (whether taken as a variable in its own-right, or seen as the time derivative of the Yang-Mills potential). This is also compatible with loop quantum cosmology and effective LQG Reissner-Nordstr\"{o}m studies where the quantum effects of matter are not manifest explicitly, but in their coupling to the effective theory via the quantum corrected equations of motion.

In figures \ref{fig:1_tabular} and \ref{fig:2_tabular} we illustrate several evolutions for initial conditions of EYM black holes which do not possess large metric fluctuations in their purely classical counter-parts \cite{ref:donets}. From these solutions one may see the quantum bounce in fig. (a) where $\pc$ does not go to zero for the holonomy corrected evolution. On the left side of the evolution (the ``classically forbidden'' region) $\pc$ oscillates in a damped manner. This is reminiscent of the behavior in holonomy corrected Schwarzschild vacuum black holes. The Yang-Mills potential, $\potb$, is illustrated in figures (b). In figures (c) and (d) the electric field component is shown. As the corrected evolution progresses to the left one can see that $\elecb$'s magnitude grows monotonically. Admittedly this is the coordinate value of the electric field, $\elecb=-\mathscr{E}_{1}^{\phi}$, but one may project this into the orthonormal frame, which yields $\hat{\elecb}:=-\mathscr{E}_{1}^{\hat{\phi}}=-\mathscr{E}_{1}^{\phi} T \sin\theta$, whose magnitude grows even more drastically\footnote{In the non-Abelian case a $\phi$ component of an electric field is not incompatible with spherical symmetry \cite{ref:miyachi}.}. One effect on the geometry is the large growth in the $R$ subspace as the evolution progresses. This may be seen in figures (e), where the proper length scale factor for the $X$ direction is illustrated, $\sqrt{\qxx}$. This scale factor grows monotonically as the evolution progresses (becoming approximately linear asymptotically). In (f) the lapse function is shown to asymptote to a constant. Therefore, asymptotically on the far side of the bounce, the geometry approaches that of the following line element:
\begin{equation}
\Scale[0.83]{ \mathrm{d}s^{2}=-N_{\mbox{\tiny{0}}}^{2}\,\mathrm{d}T^{2} + \alpha_{\mbox{\tiny{0}}}T^{n}\, \mathrm{d}X^{2} + \beta{\mbox{\tiny{0}}}^{2}\left(\mathrm{d}\theta^{2} + \sin^{2}\theta\, \mathrm{d}\phi^{2}\right),} \label{eq:asymptline}
\end{equation}
where the subscript {\tiny{0}} indicates a constant quantity. It is interesting to note that, at least qualitatively, this is not unlike the behavior of a loop quantum corrected pure vacuum black hole \cite{ref:modestobhsing2}, \cite{ref:BandV}, \cite{ref:modestobhsing}, as well as those with cosmological constant and non-trivial topologies \cite{ref:BKD}, indicating that there may be some sort of uniqueness theorem in the long-time limit for holonomy corrected black holes. In all cases there is a quantum bounce, and the qualitative behavior on the other side of the bounce in the EYM scenarios studied here mimic the behavior in \cite{ref:BandV} and \cite{ref:BKD}, both which utilized the $\overline{\mu}^{\prime}$ scheme. That is, $\pc$ tends to a constant on the far side of the bounce in the long-time evolution. It was later shown that the resulting spacetime on the other side of the bounce for the spherically symmetric black holes and higher genus black holes can be interpreted as Nariai and Bertotti-Robinson type with properties quite different from the pre-bounce conditions \cite{ref:DJS}.

\begin{figure}[!htp]
\begin{framed}
\begin{center}
\vspace{-0.5cm}
\begin{tabular}{cc}
\subfloat[\footnotesize{The two-sphere scale factor $\pc$.}]{\includegraphics[width=0.50\textwidth, height=0.48\textwidth, clip,viewport=66 360 435 760]{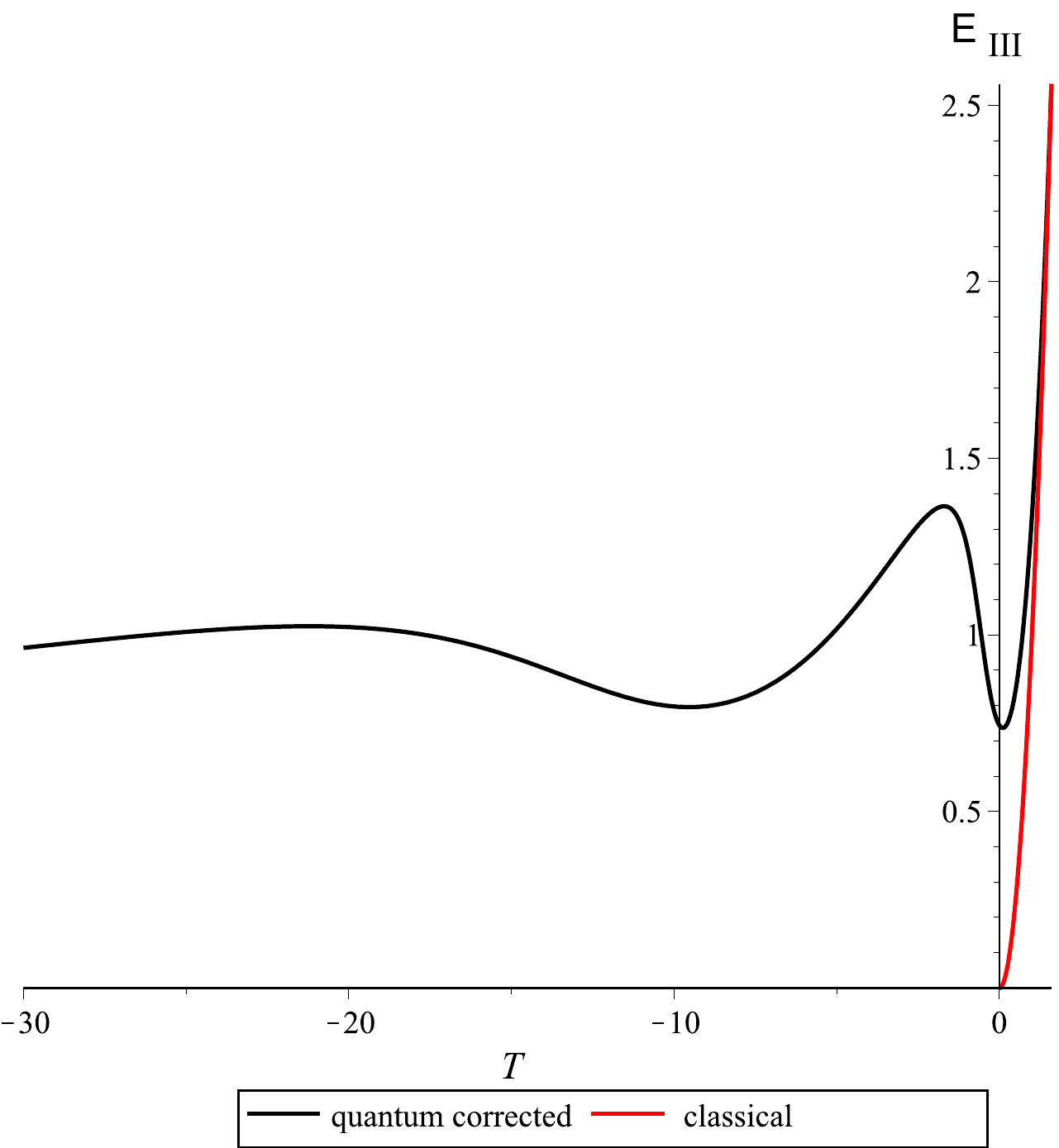}}&\hspace{0.5cm}
\subfloat[\footnotesize{The Yang-Mills potential $\potb$.}]{\includegraphics[width=0.50\textwidth, height=0.45\textwidth, clip,viewport=65 438 435 730]{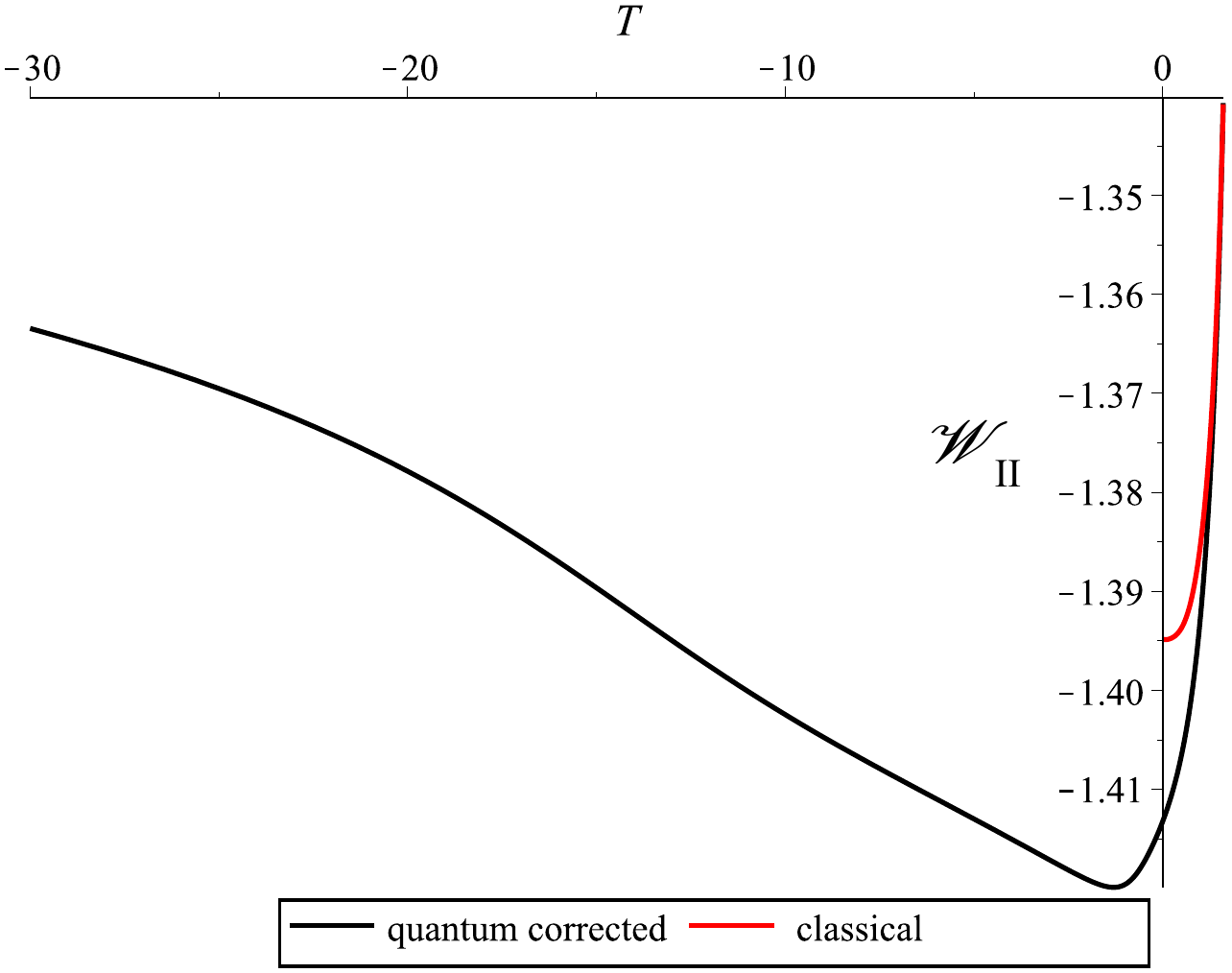}} \\
\subfloat[\footnotesize{The Yang-Mills electric field $\elecb$ near the classical singular point.}]{\includegraphics[width=0.50\textwidth, height=0.45\textwidth, clip,viewport=90 438 435 735]{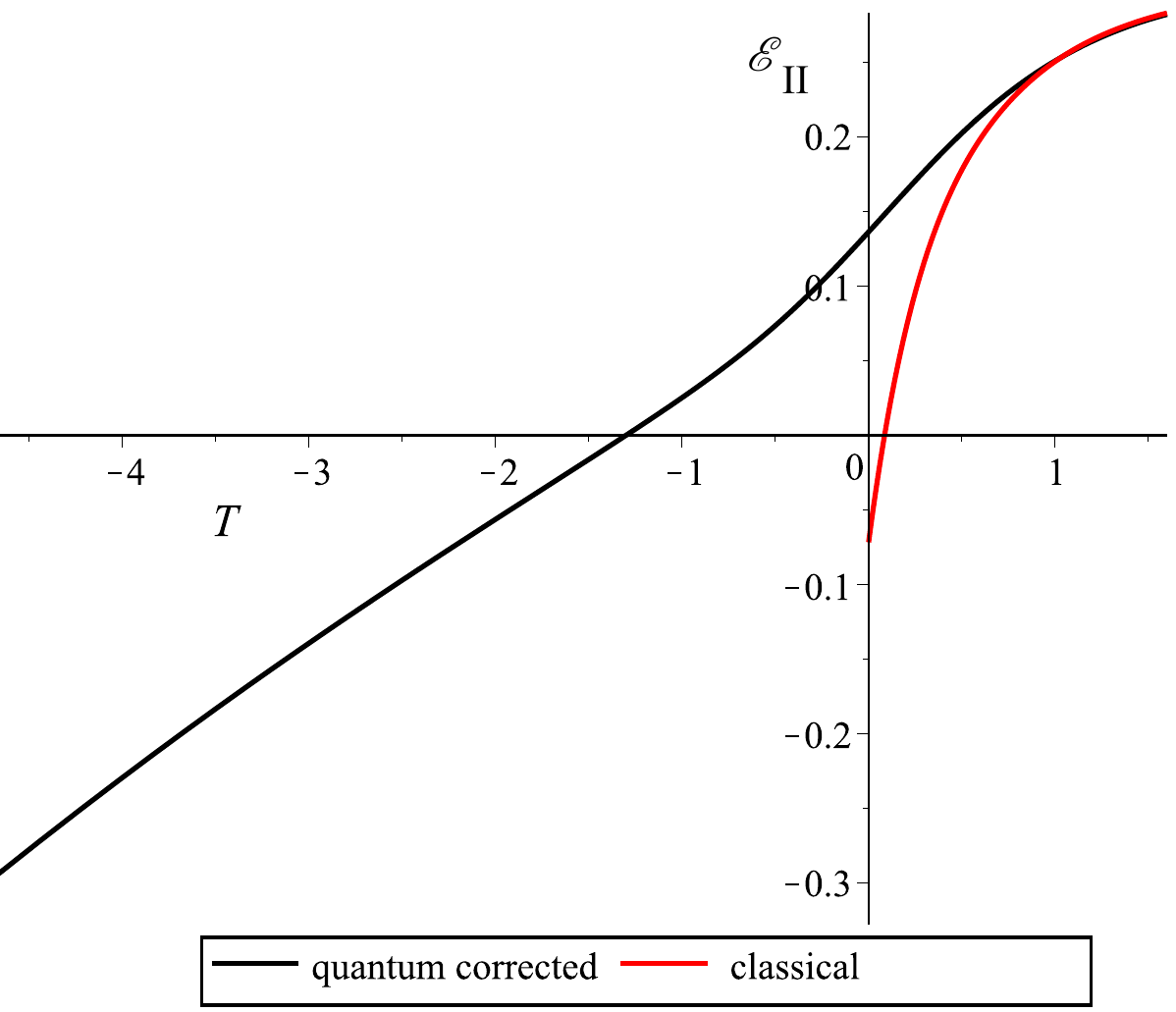}}&\hspace{0.5cm}
\subfloat[\footnotesize{The long time evolution of the Yang-Mills electric field $\elecb$.}]{\includegraphics[width=0.50\textwidth, height=0.45\textwidth, clip,viewport=65 440 435 730]{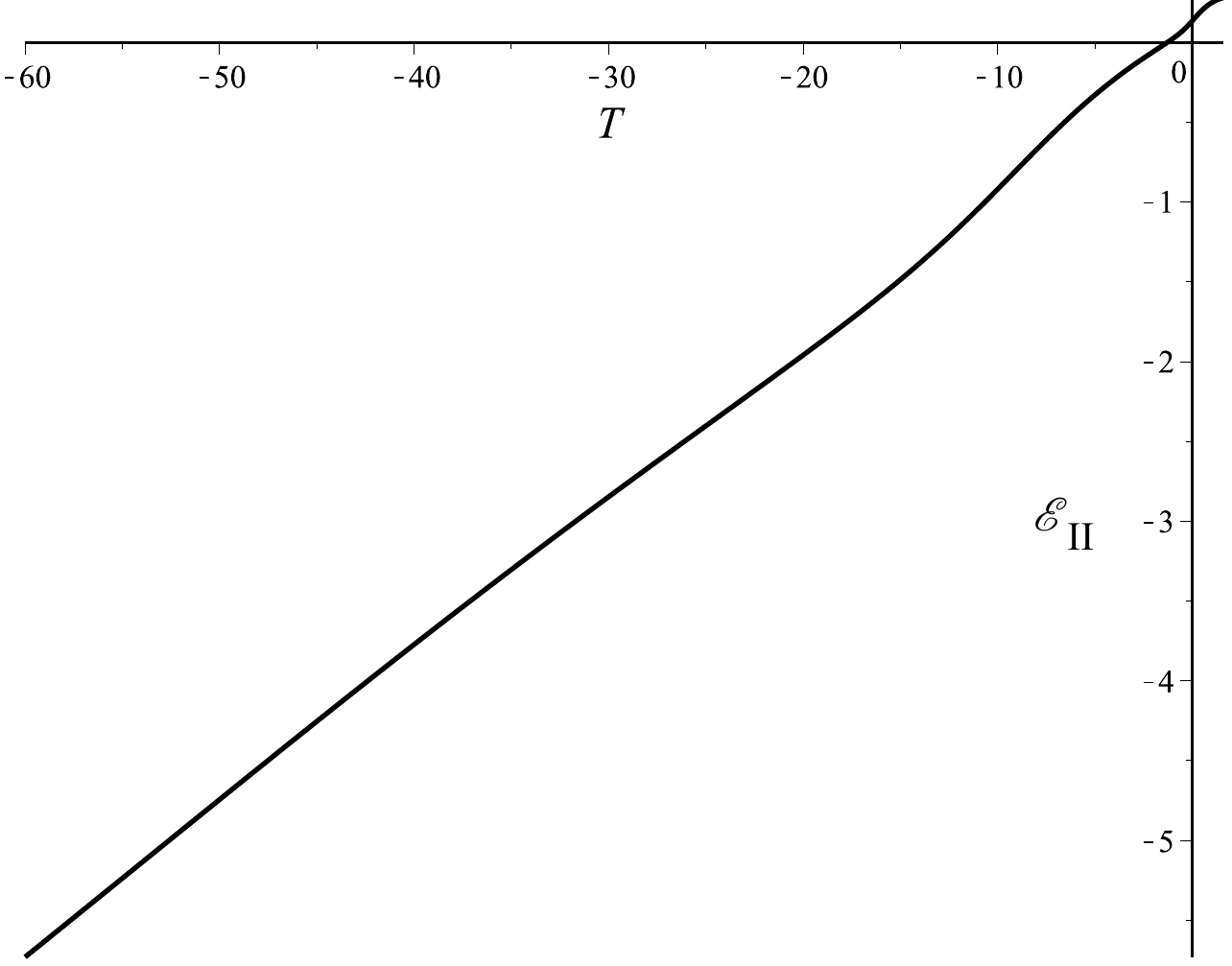}}\\
\subfloat[\footnotesize{The scale factor $\sqrt{\qxx}$.}]{\includegraphics[width=0.50\textwidth, height=0.48\textwidth, clip,viewport=65 445 475 735]{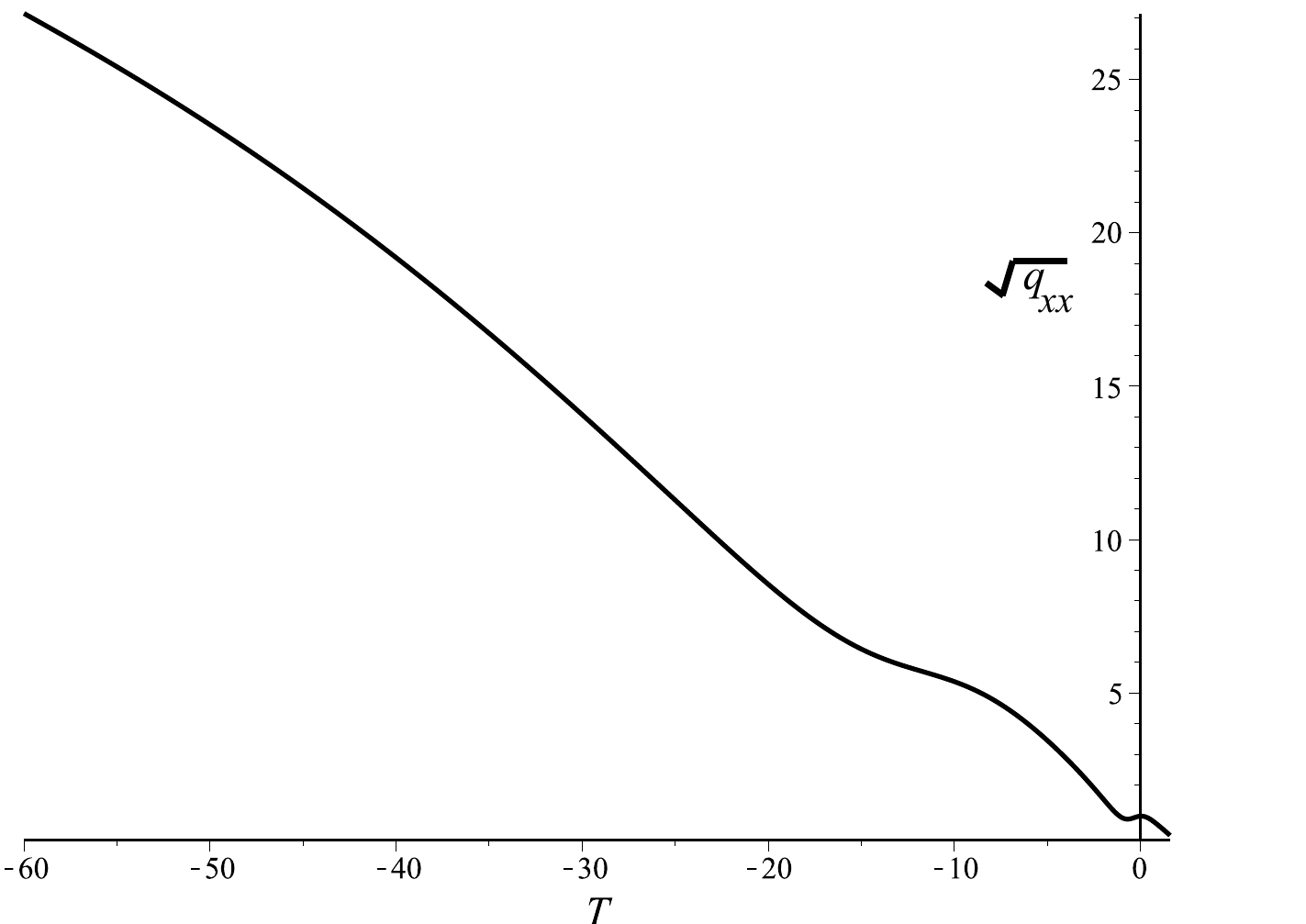}}&\hspace{0.5cm}
\subfloat[\footnotesize{The lapse function $N$.}]{\includegraphics[width=0.50\textwidth, height=0.45\textwidth, clip,viewport=65 445 435 735]{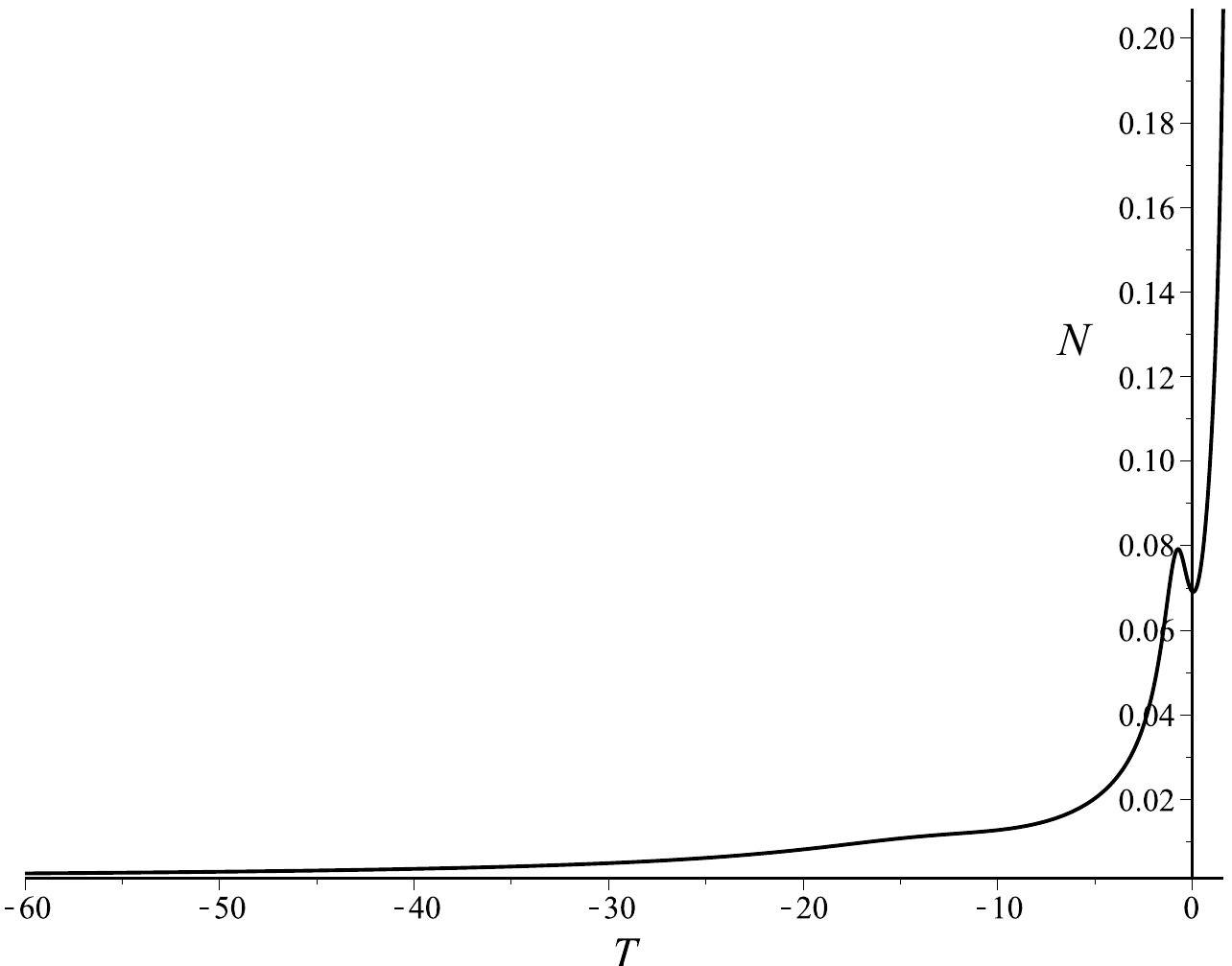}}
\end{tabular}
\end{center}
\caption{\small{An evolution of the two-sphere scale factor ($\pc$), the Yang-Mills potential ($\potb$), and the Yang-Mills electric field ($\elecb$) for both classical and holonomy corrected evolution equations. In the quantum corrected case the volume of two-spheres does not shrink to zero, as may be seen in fig. (a), indicating the quantum bounce. Gravitational holonomy corrections only.}}
\label{fig:1_tabular}
\end{framed}
\end{figure}

\begin{figure}[!htp]
\begin{framed}
\begin{center}
\vspace{-0.5cm}
\begin{tabular}{cc}
\subfloat[\footnotesize{The two-sphere scale factor $\pc$.}]{\includegraphics[width=0.50\textwidth, height=0.48\textwidth, clip,viewport=70 438 435 756]{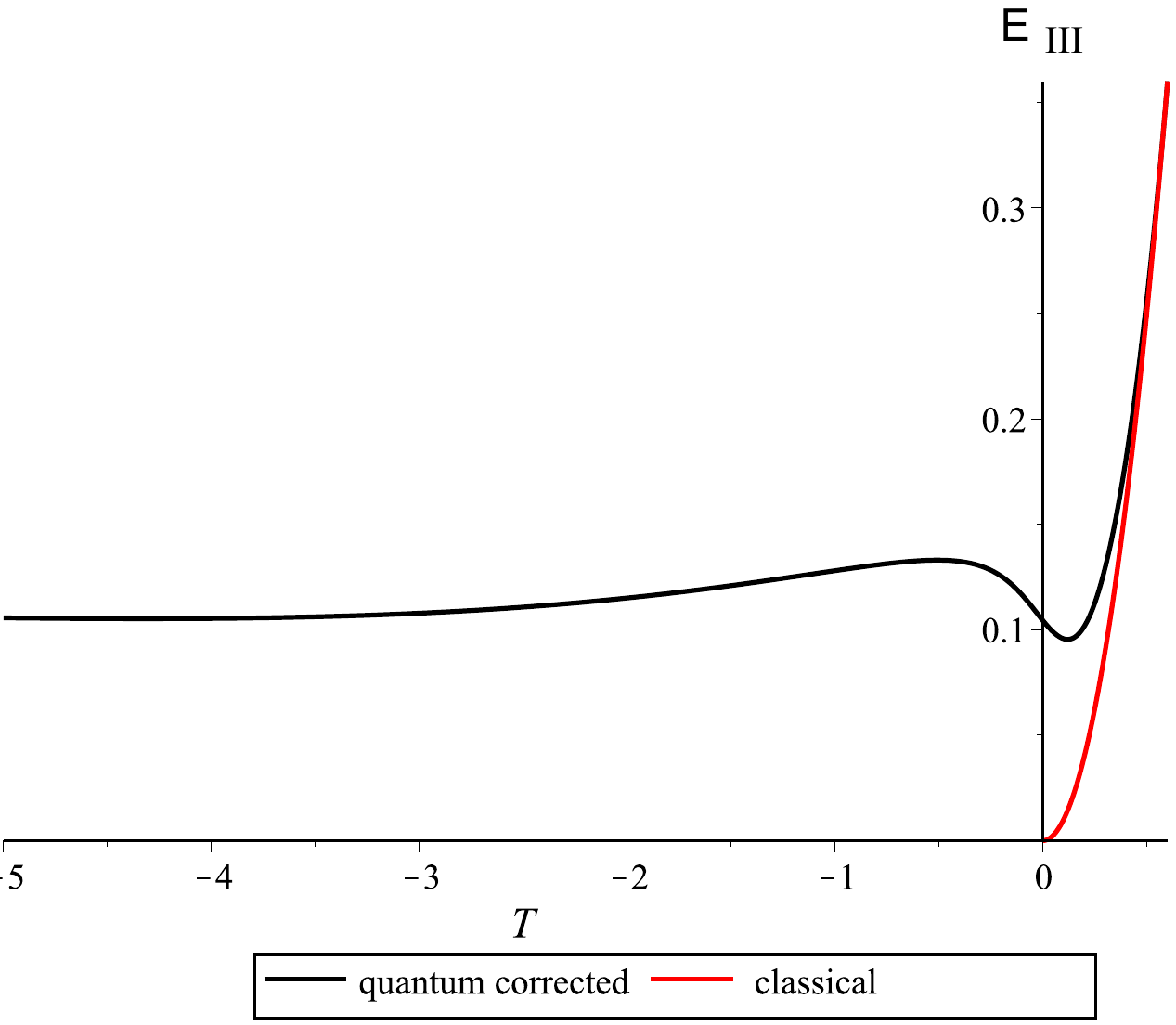}}&\hspace{0.5cm}
\subfloat[\footnotesize{The Yang-Mills potential $\potb$.}]{\includegraphics[width=0.50\textwidth, height=0.45\textwidth, clip,viewport=65 420 435 730]{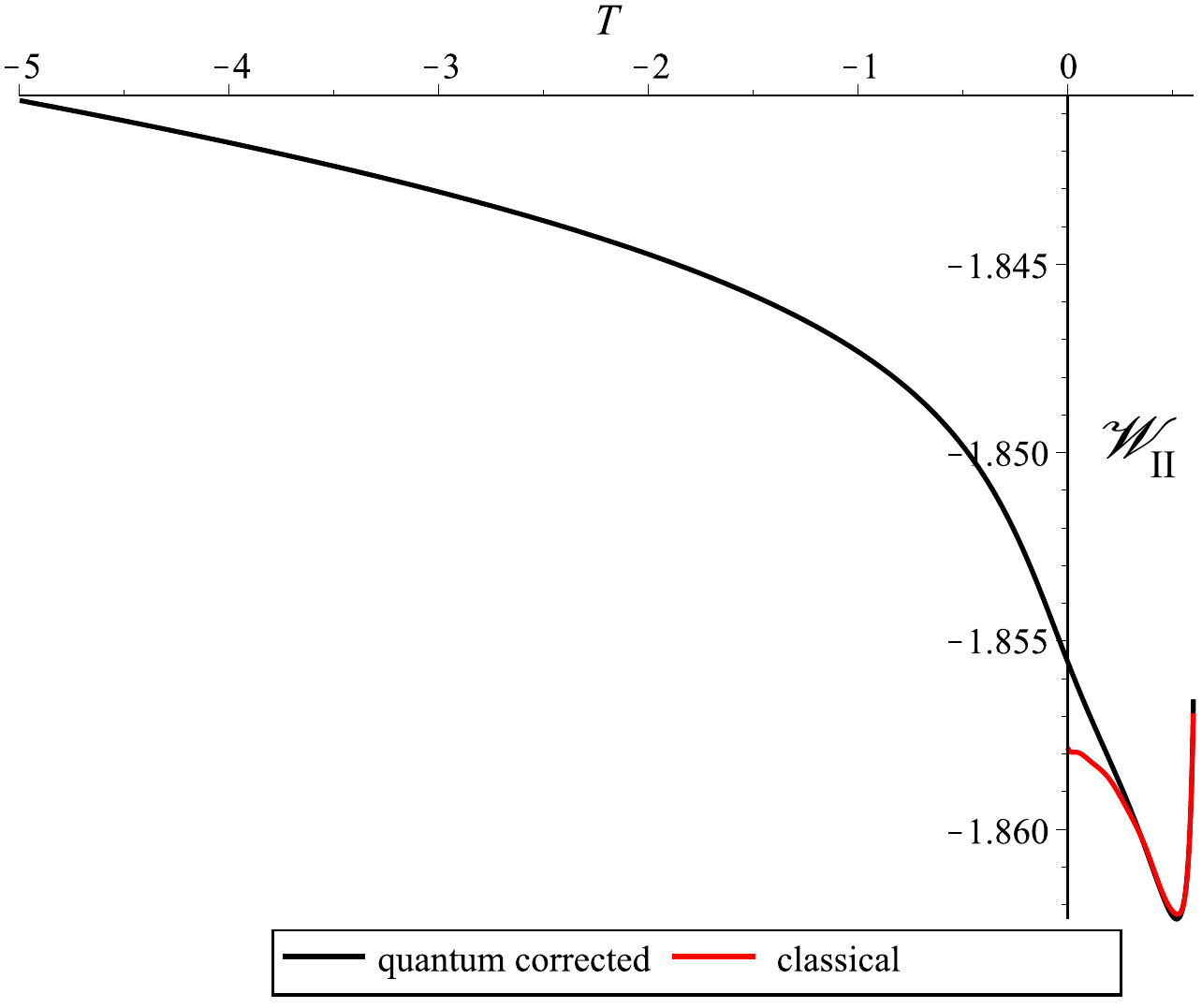}} \\
\subfloat[\footnotesize{The Yang-Mills electric field $\elecb$ near the classical singular point.}]{\includegraphics[width=0.50\textwidth, height=0.45\textwidth, clip,viewport=90 420 435 735]{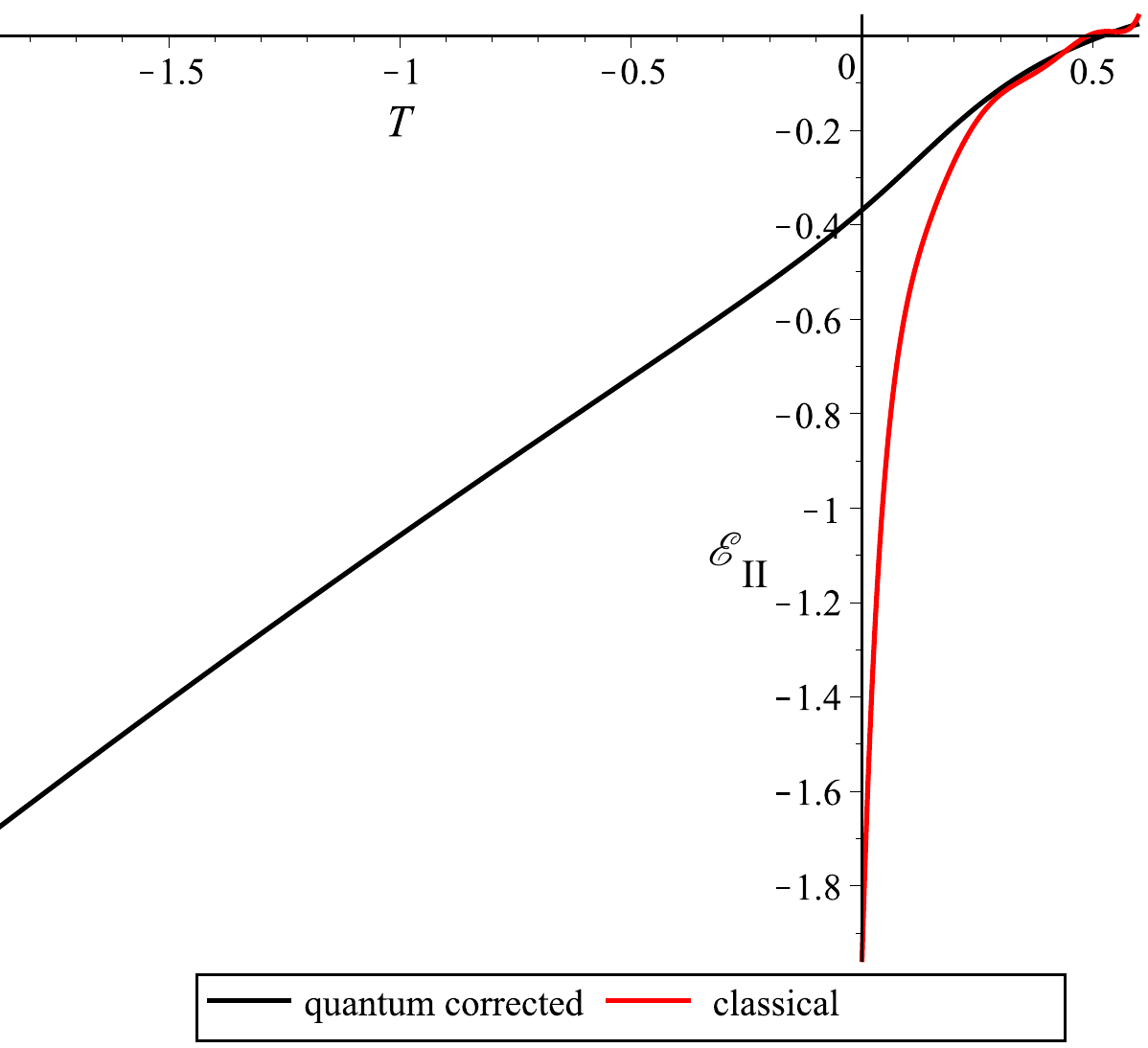}}&\hspace{0.5cm}
\subfloat[\footnotesize{The long time evolution of the Yang-Mills electric field $\elecb$.}]{\includegraphics[width=0.50\textwidth, height=0.45\textwidth, clip,viewport=65 440 435 730]{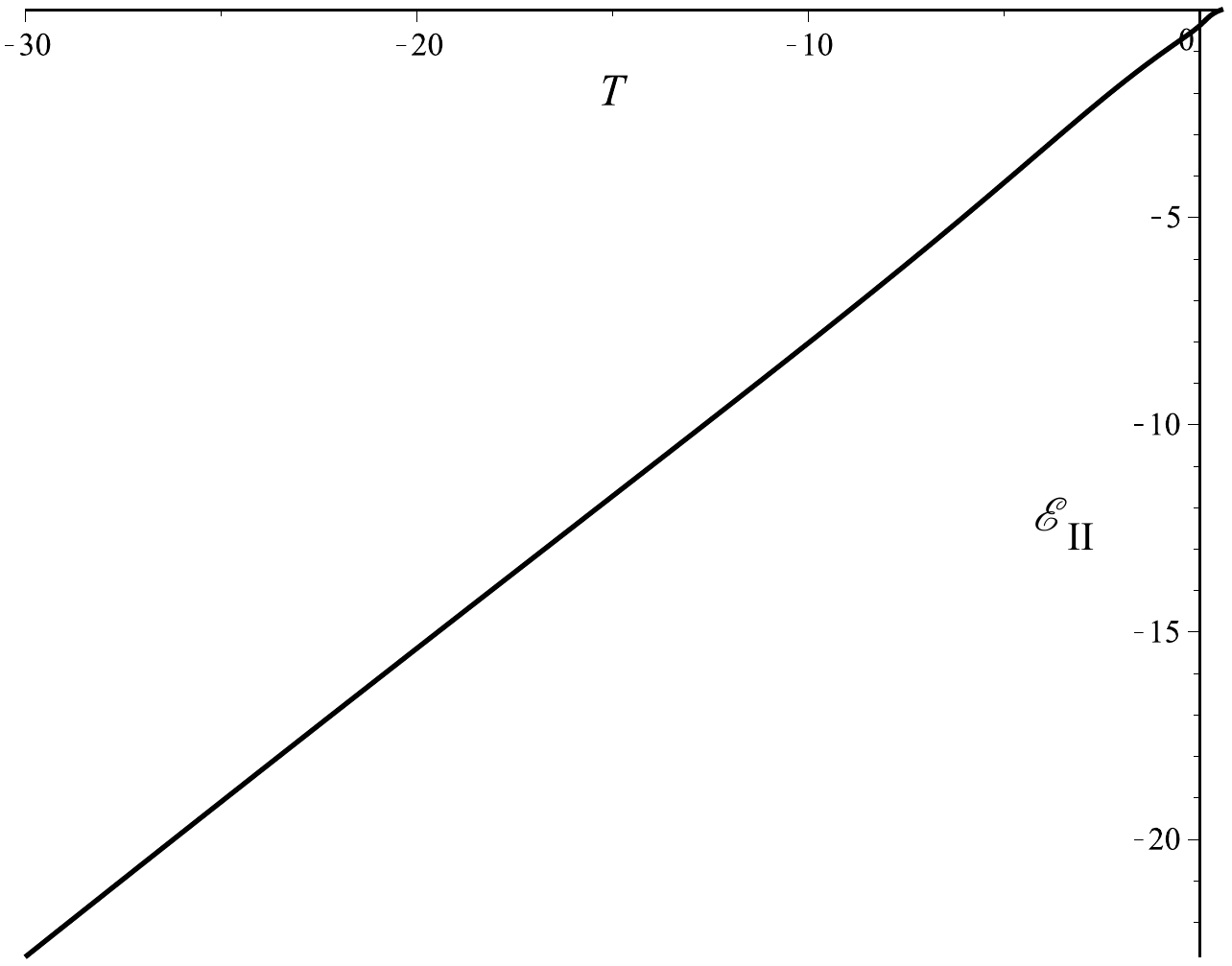}}\\
\subfloat[\footnotesize{The scale factor $\sqrt{\qxx}$.}]{\includegraphics[width=0.50\textwidth, height=0.48\textwidth, clip,viewport=65 445 475 735]{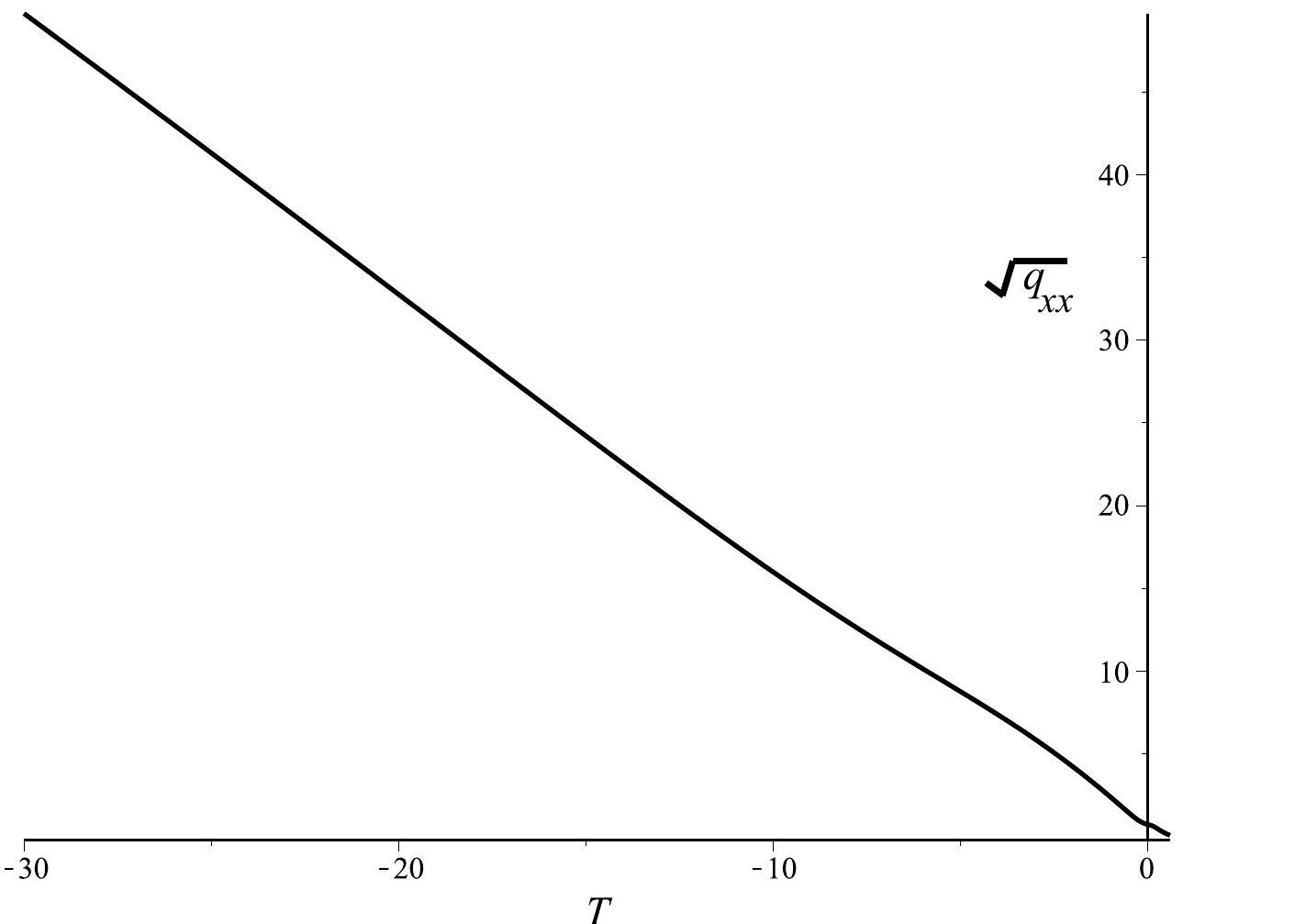}}&\hspace{0.5cm}
\subfloat[\footnotesize{The lapse function $N$.}]{\includegraphics[width=0.50\textwidth, height=0.45\textwidth, clip,viewport=65 445 435 735]{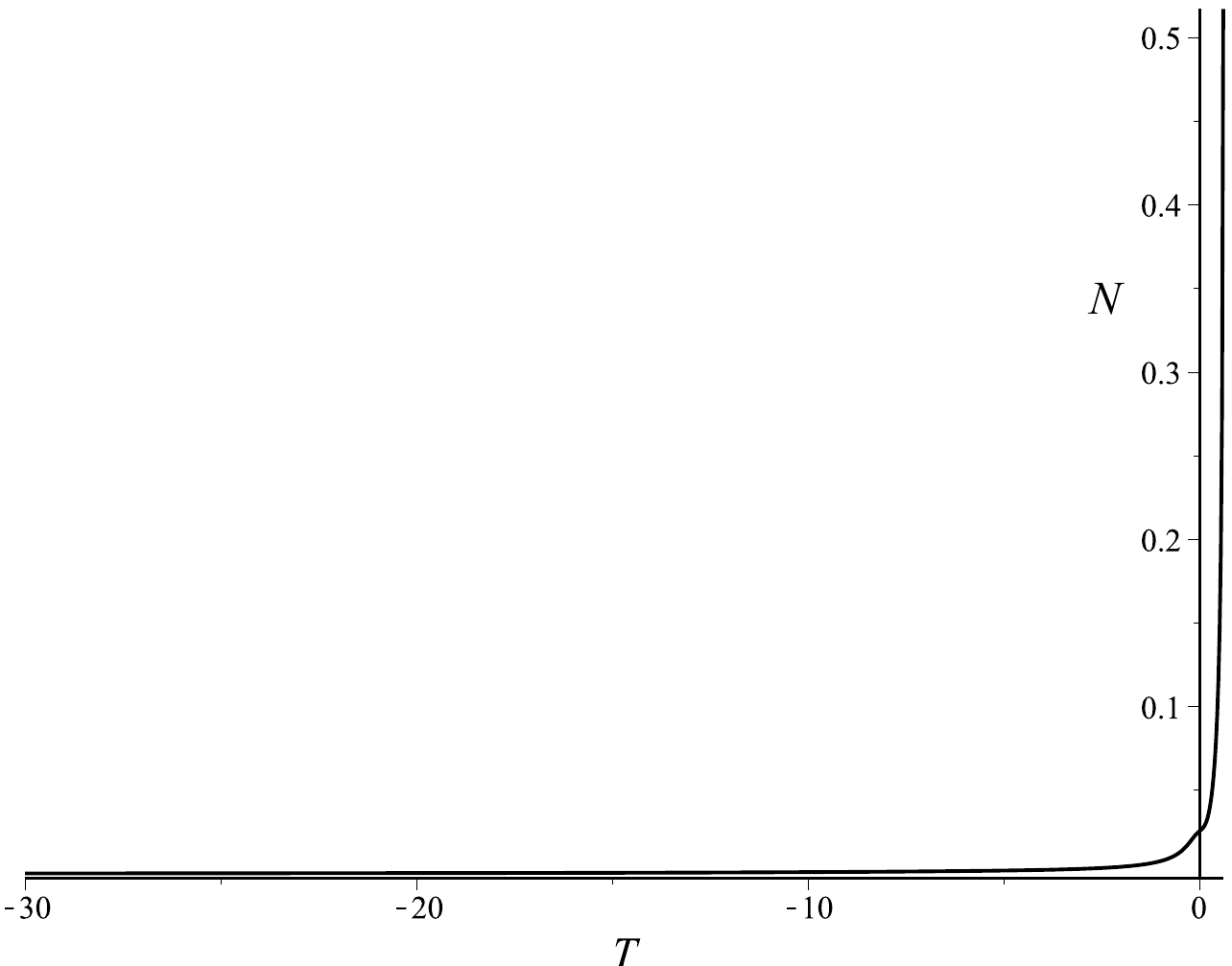}}
\end{tabular}
\end{center}
\caption{\small{Another evolution of the two-sphere scale factor ($\pc$), the Yang-Mills potential ($\potb$), and the Yang-Mills electric field ($\elecb$) for both classical and holonomy corrected evolution equations. In the quantum corrected case the volume of two-spheres does not shrink to zero, as may be seen in fig. (a), indicating the quantum bounce. Gravitational holonomy corrections only.}}
\label{fig:2_tabular}
\end{framed}
\end{figure}

Next we consider the same evolutions as above but in the case where we implement Wilson loop corrections on the Yang-Mills field. The rationale for this is that the correction argument, as outlined in the appendix, can just as easily be applied to the Yang-Mills field strength as the gravitational one. This will implement some low-order ``quantum'' corrections due to the field potential propagating on a lattice structure due to the discrete area/volume structure of loop quantum gravity. These corrections are summarized as:
\begin{subequations}
\romansubs
{\allowdisplaybreaks\begin{align}
\potb \rightarrow& \frac{\sin\left(\potb \delta\subsX\right)}{\delta\subsX}\,, \label{eq:wilsonb}\\
\potc \rightarrow& \frac{\sin\left(\potc \delta_{\theta}\right)}{\delta_{\theta}}\,. \label{eq:wilsonc}
\end{align}}
\end{subequations}
It should be noted though that the corrections (\ref{eq:wilsonb}-ii) for non-zero $\delta$ may interfere with the $SU(2)$ gauge invariance of the Yang-Mills equations of motion. We show the resulting evolutions in figures \ref{fig:1_tabular_YM} and \ref{fig:2_tabular_YM}.

\begin{figure}[!htp]
\begin{framed}
\begin{center}
\vspace{-0.5cm}
\begin{tabular}{cc}
\subfloat[\footnotesize{The two-sphere scale factor $\pc$.}]{\includegraphics[width=0.50\textwidth, height=0.48\textwidth, clip,viewport=65 438 435 755]{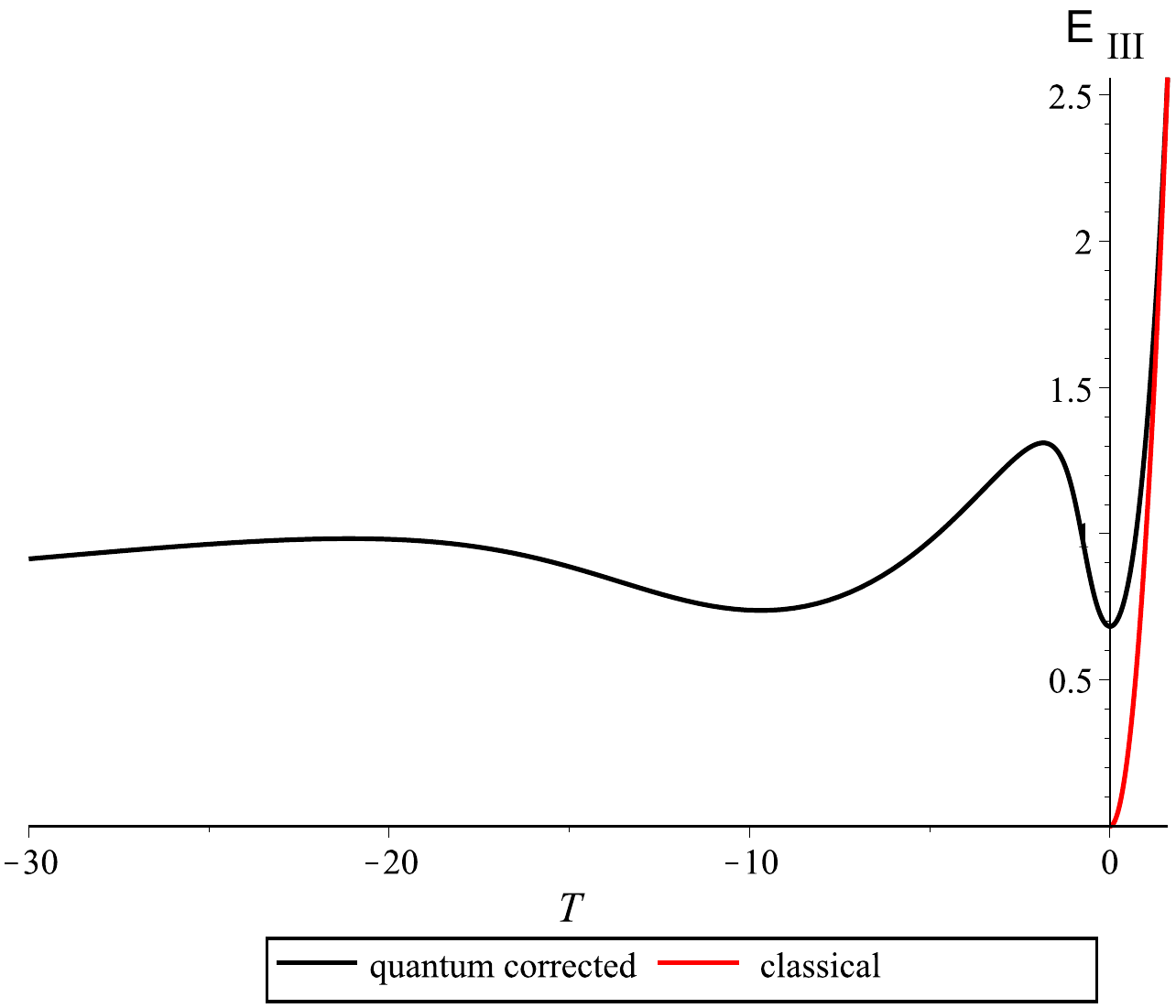}}&\hspace{0.5cm}
\subfloat[\footnotesize{The Yang-Mills potential $\potb$.}]{\includegraphics[width=0.50\textwidth, height=0.45\textwidth, clip,viewport=65 438 435 730]{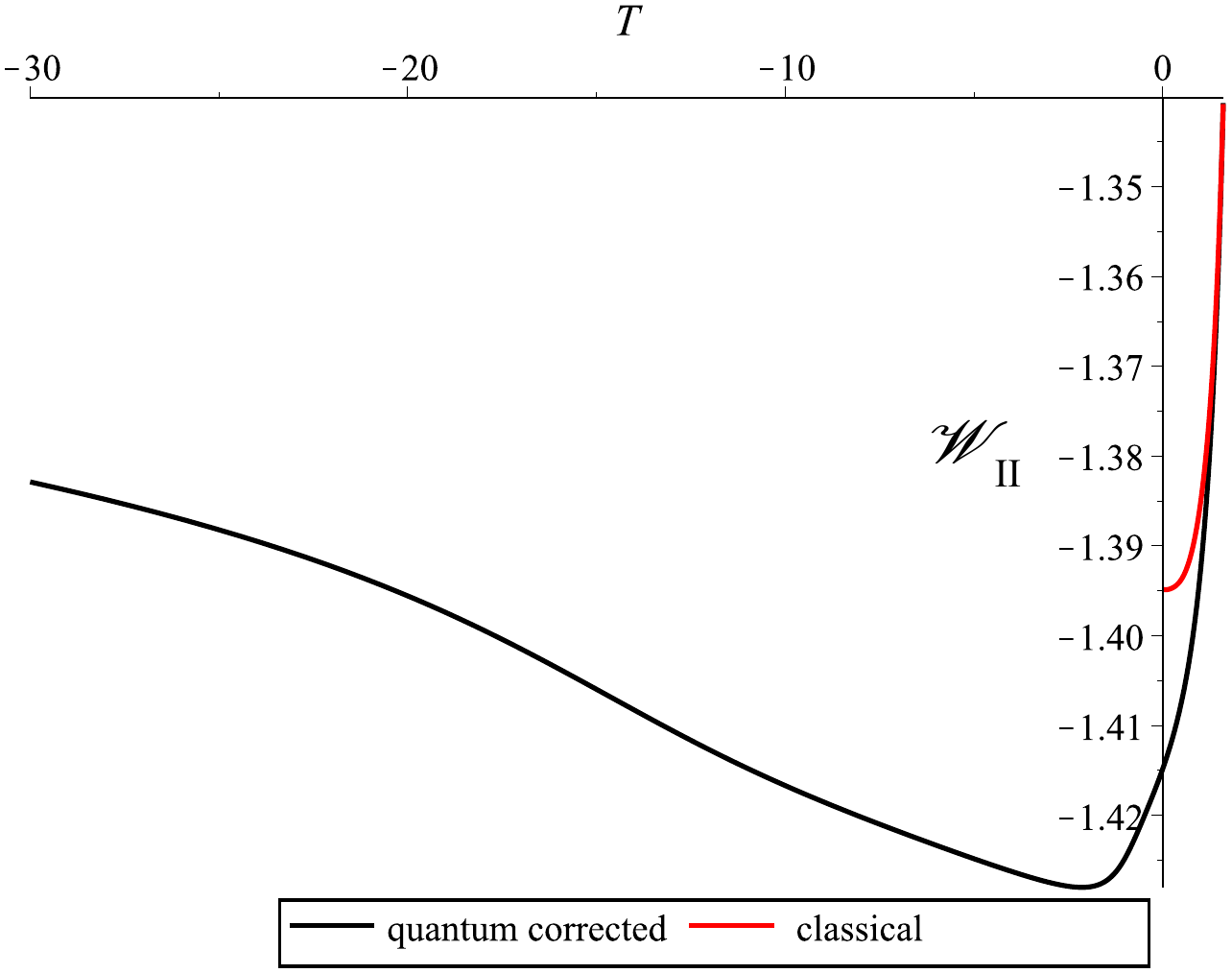}} \\
\subfloat[\footnotesize{The Yang-Mills electric field $\elecb$ near the classical singular point.}]{\includegraphics[width=0.50\textwidth, height=0.45\textwidth, clip,viewport=65 440 435 735]{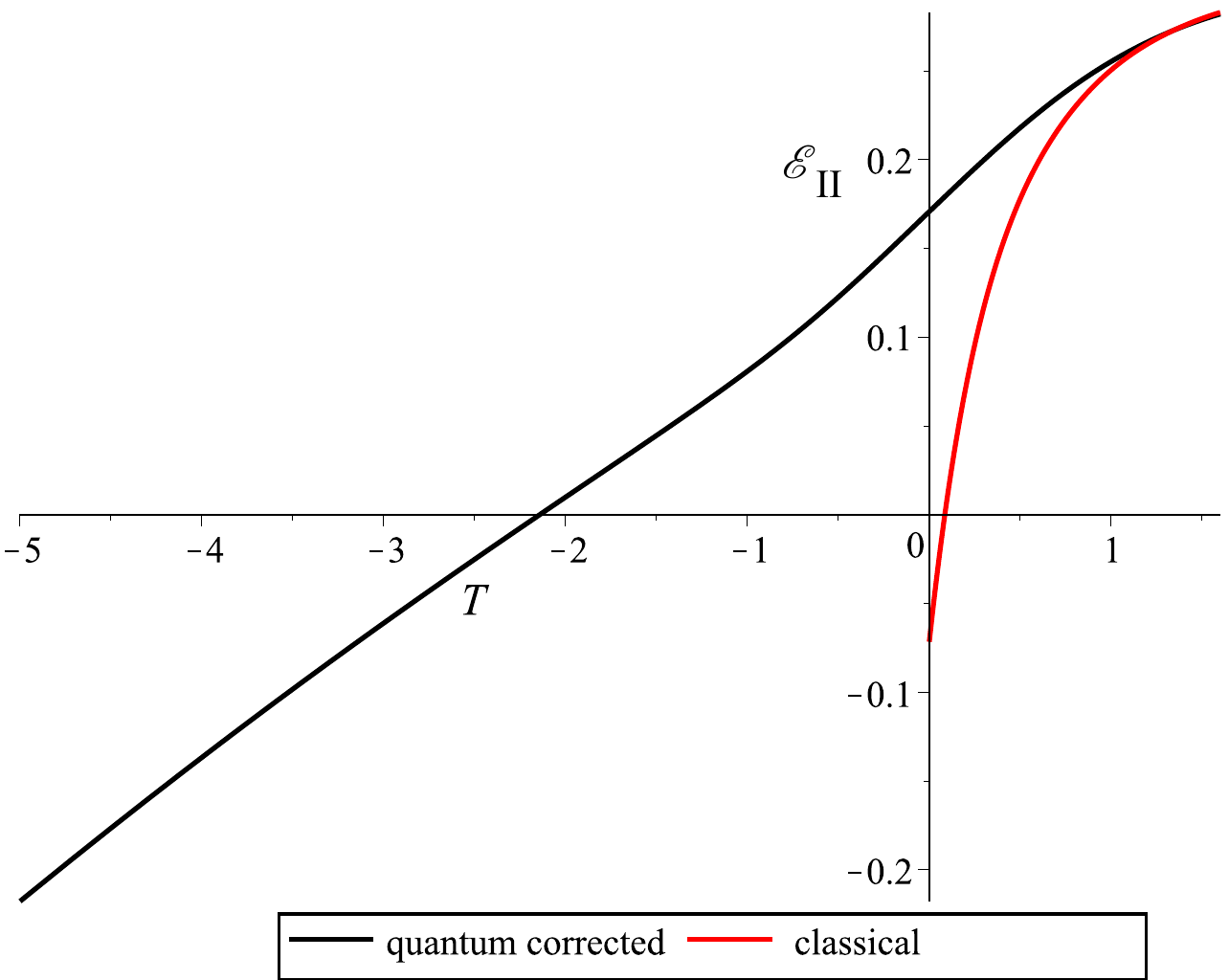}}&\hspace{0.5cm}
\subfloat[\footnotesize{The long time evolution of the Yang-Mills electric field $\elecb$.}]{\includegraphics[width=0.50\textwidth, height=0.45\textwidth, clip,viewport=70 440 435 736]{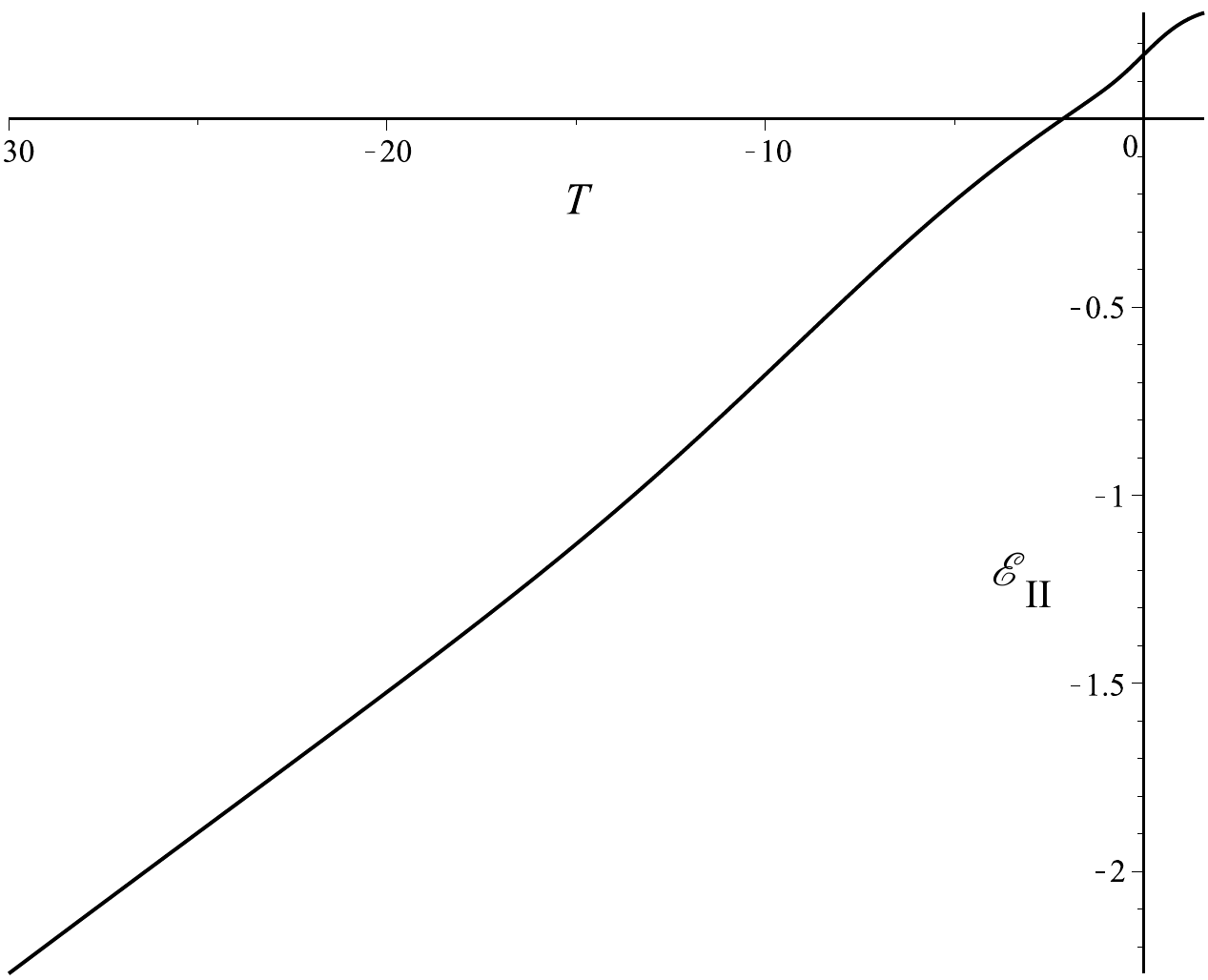}}\\
\subfloat[\footnotesize{The scale factor $\sqrt{\qxx}$.}]{\includegraphics[width=0.50\textwidth, height=0.48\textwidth, clip,viewport=65 445 435 735]{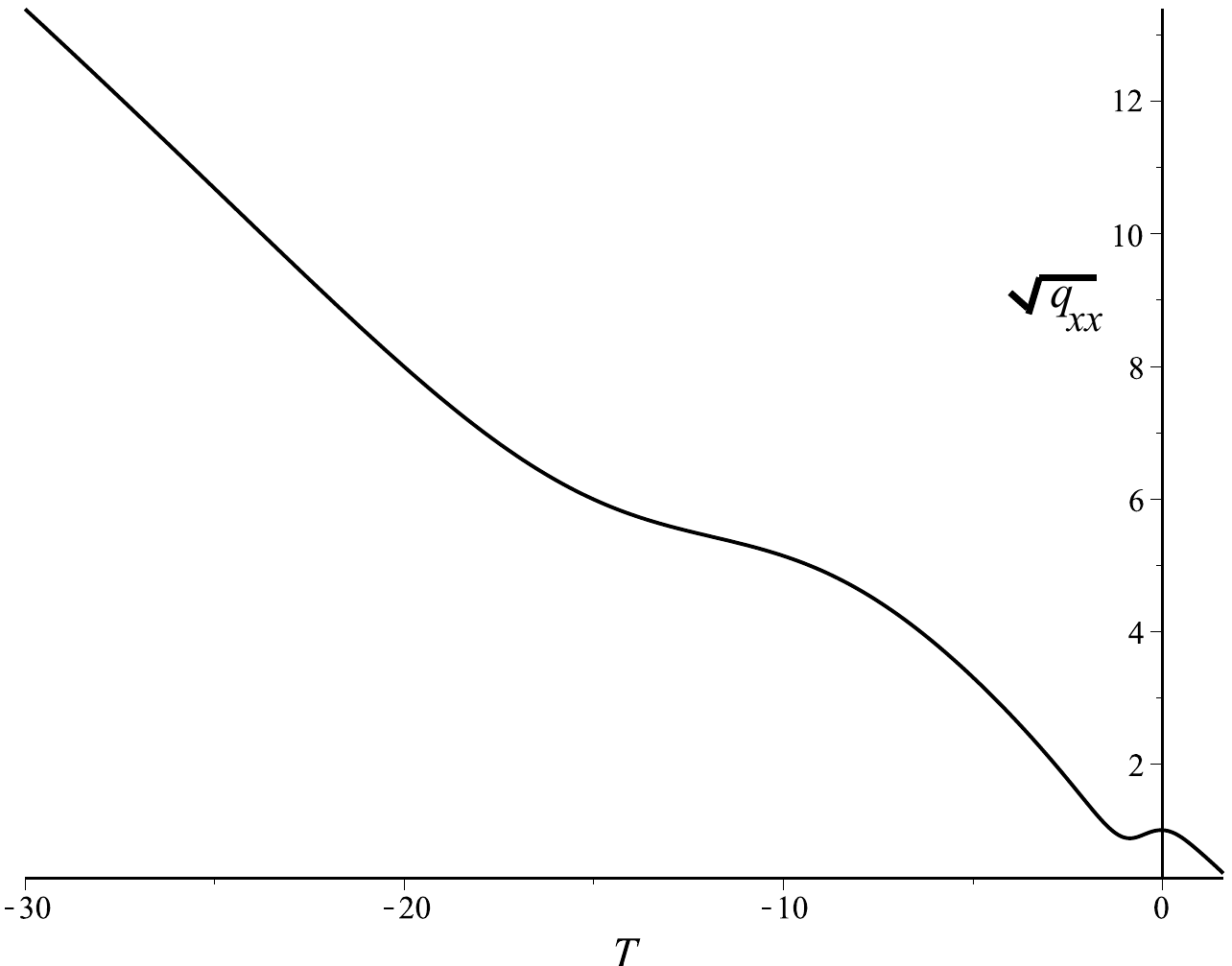}}&\hspace{0.5cm}
\subfloat[\footnotesize{The lapse function $N$.}]{\includegraphics[width=0.50\textwidth, height=0.45\textwidth, clip,viewport=65 442 435 735]{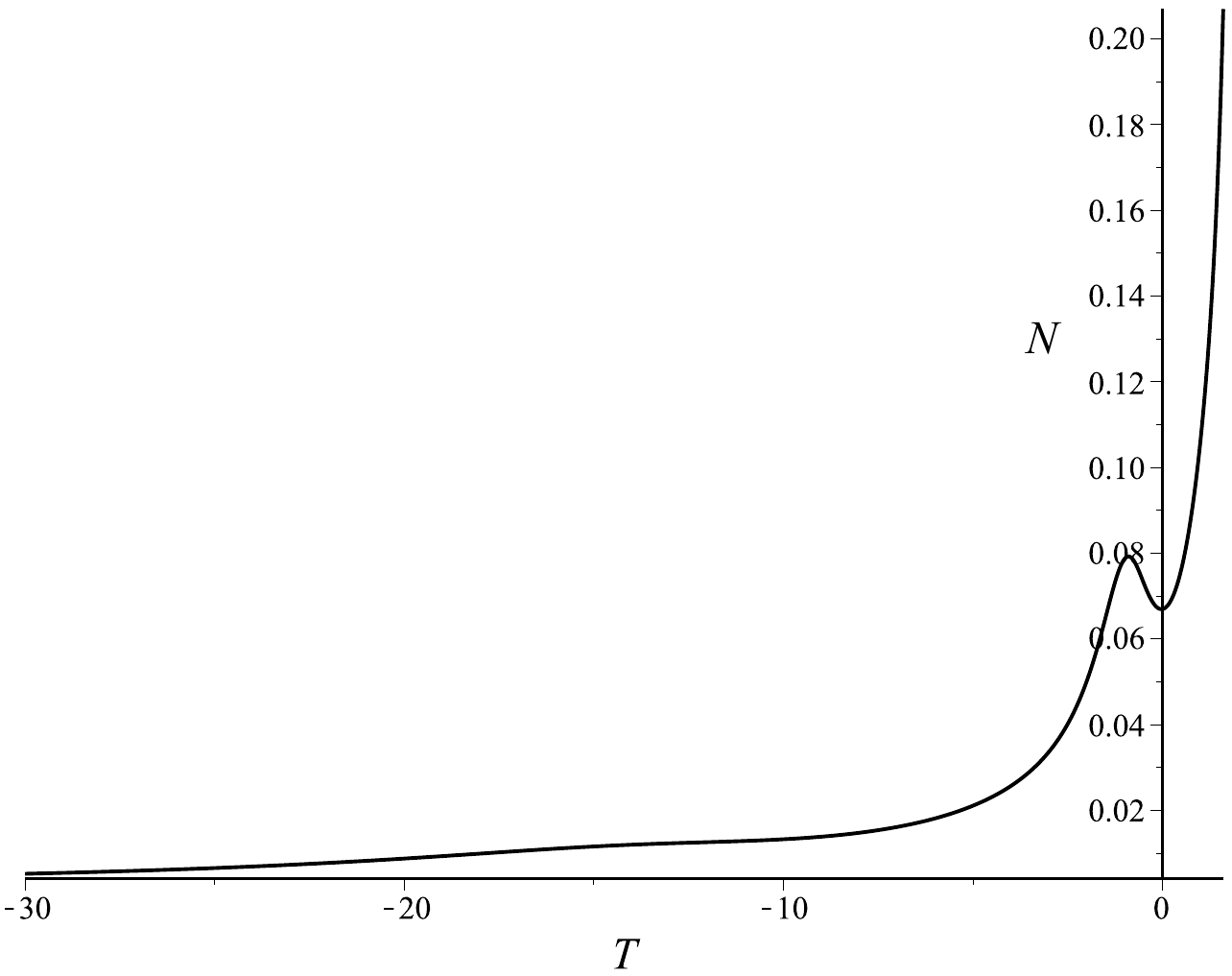}}
\end{tabular}
\end{center}
\caption{\small{An evolution of the two-sphere scale factor ($\pc$), the Yang-Mills potential ($\potb$), and the Yang-Mills electric field ($\elecb$) for both classical and holonomy corrected evolution equations. In the quantum corrected case the volume of two-spheres does not shrink to zero, as may be seen in fig. (a), indicating the quantum bounce. Gravitational holonomy corrections and Yang-Mills Wilson loop corrections.}}
\label{fig:1_tabular_YM}
\end{framed}
\end{figure}

\begin{figure}[!htp]
\begin{framed}
\begin{center}
\vspace{-0.5cm}
\begin{tabular}{cc}
\subfloat[\footnotesize{The two-sphere scale factor $\pc$.}]{\includegraphics[width=0.50\textwidth, height=0.48\textwidth, clip,viewport=65 438 435 750]{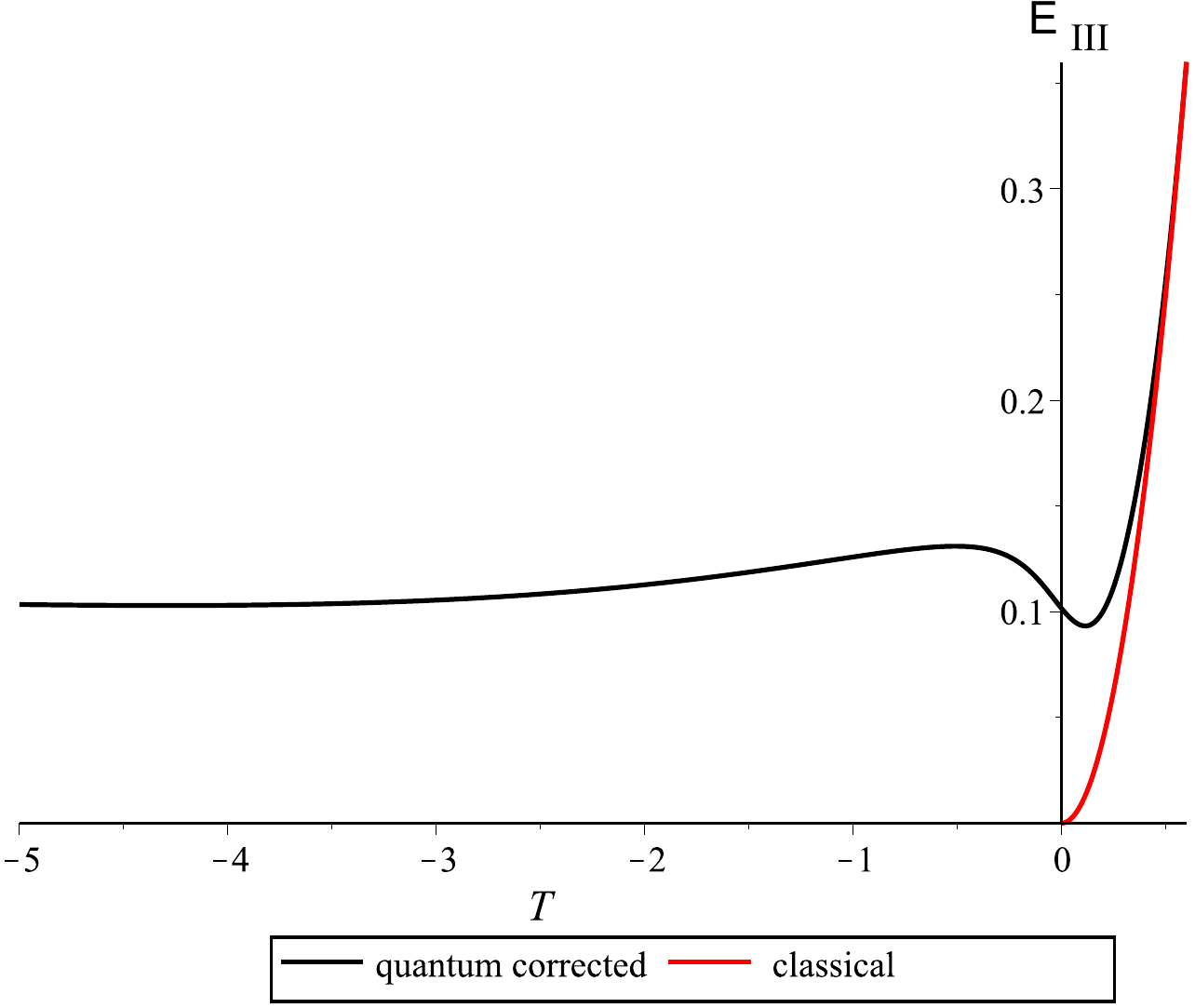}}&\hspace{0.5cm}
\subfloat[\footnotesize{The Yang-Mills potential $\potb$.}]{\includegraphics[width=0.50\textwidth, height=0.45\textwidth, clip,viewport=65 422 435 730]{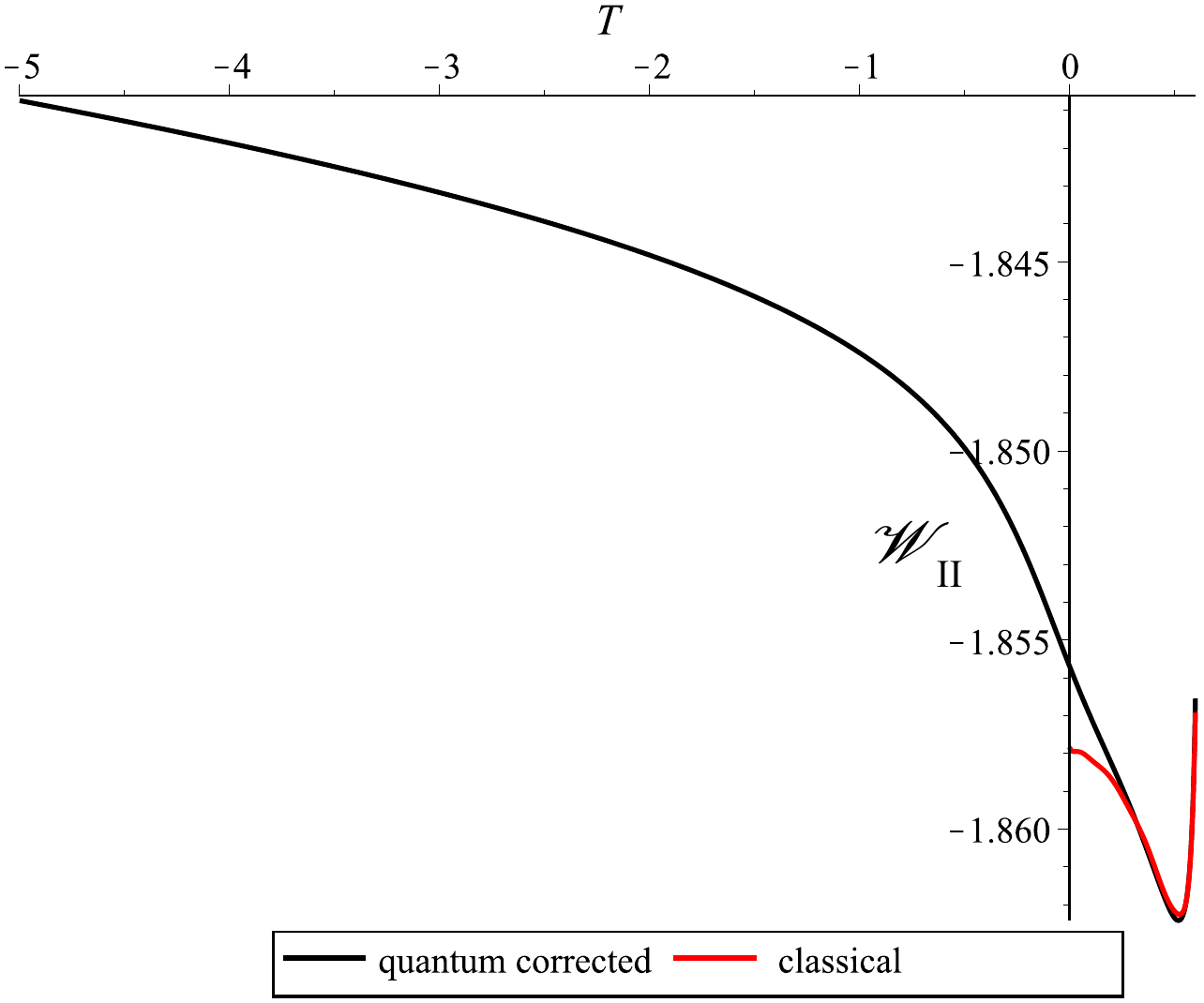}} \\
\subfloat[\footnotesize{The Yang-Mills electric field $\elecb$ near the classical singular point.}]{\includegraphics[width=0.50\textwidth, height=0.45\textwidth, clip,viewport=68 420 435 735]{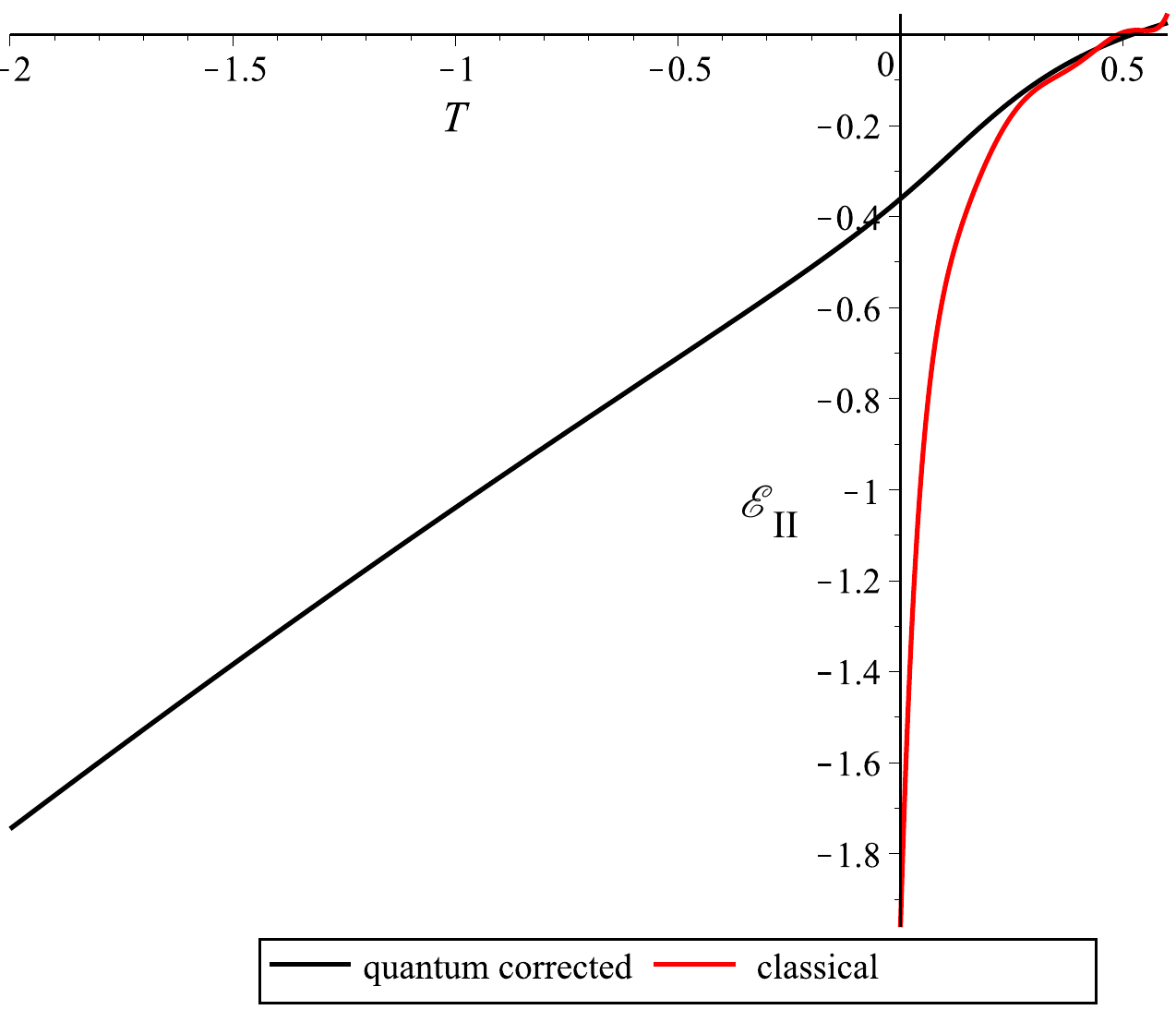}}&\hspace{0.5cm}
\subfloat[\footnotesize{The long time evolution of the Yang-Mills electric field $\elecb$.}]{\includegraphics[width=0.50\textwidth, height=0.45\textwidth, clip,viewport=65 440 435 730]{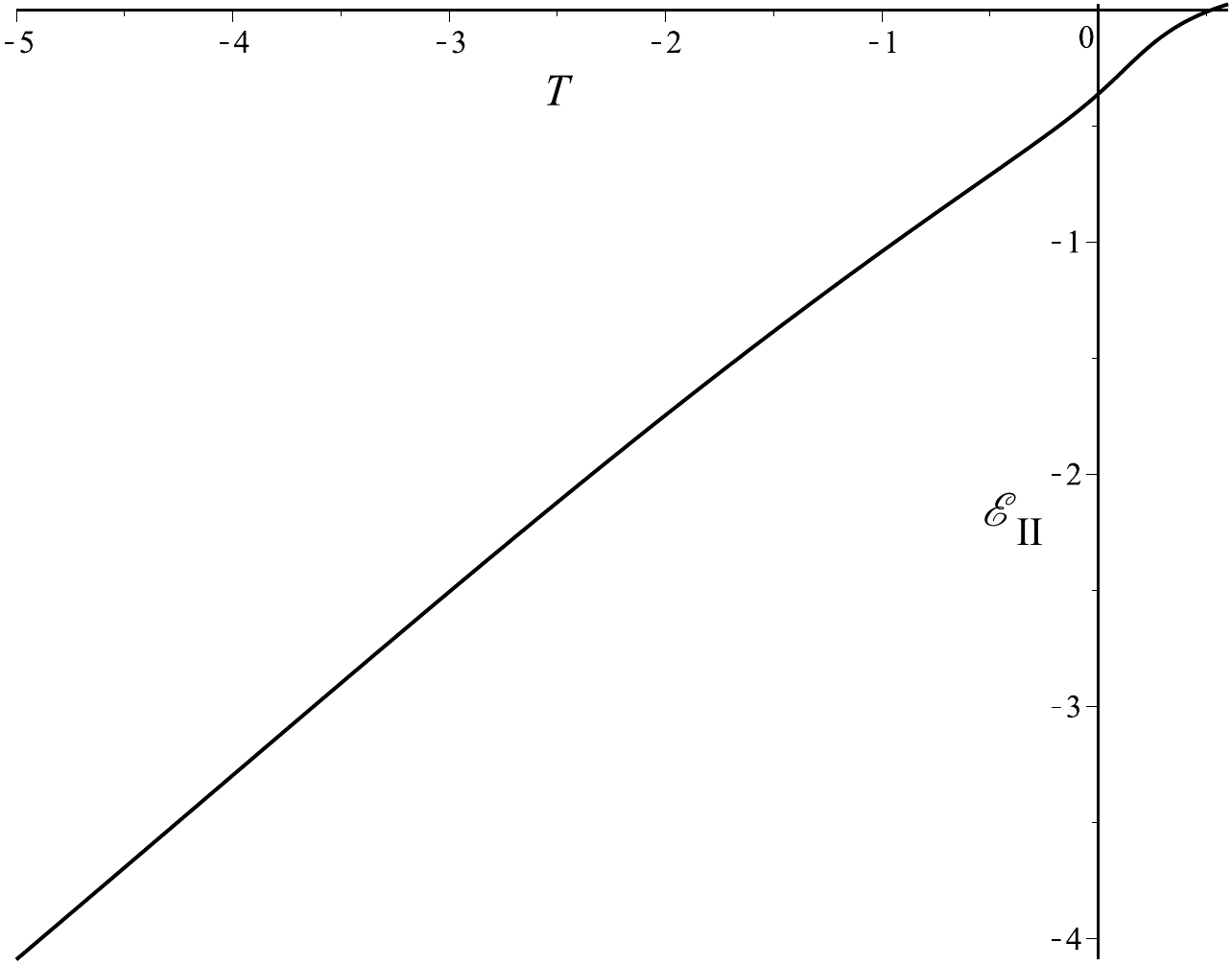}}\\
\subfloat[\footnotesize{The scale factor $\sqrt{\qxx}$.}]{\includegraphics[width=0.50\textwidth, height=0.48\textwidth, clip,viewport=65 445 435 730]{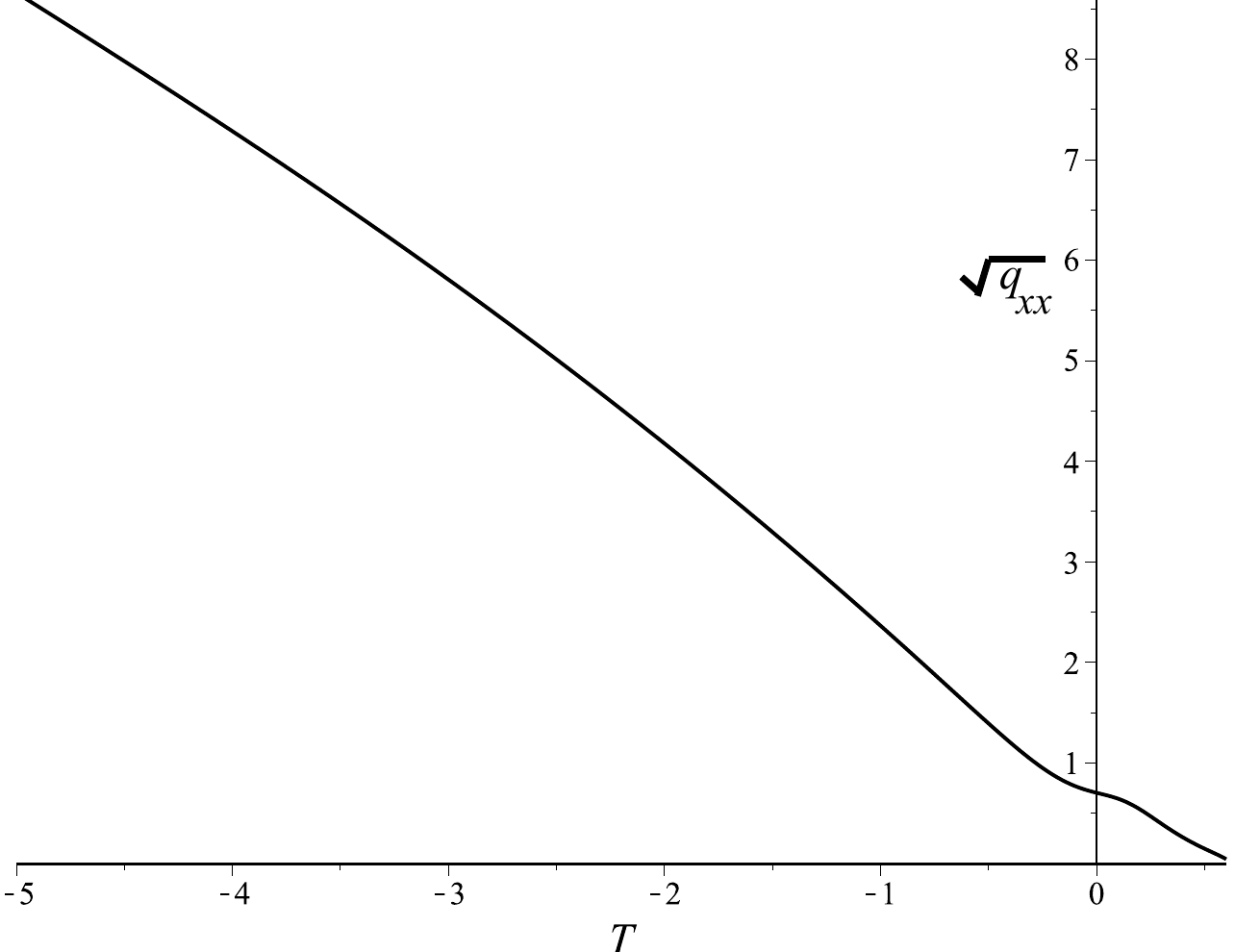}}&\hspace{0.5cm}
\subfloat[\footnotesize{The lapse function $N$.}]{\includegraphics[width=0.50\textwidth, height=0.45\textwidth, clip,viewport=65 445 435 735]{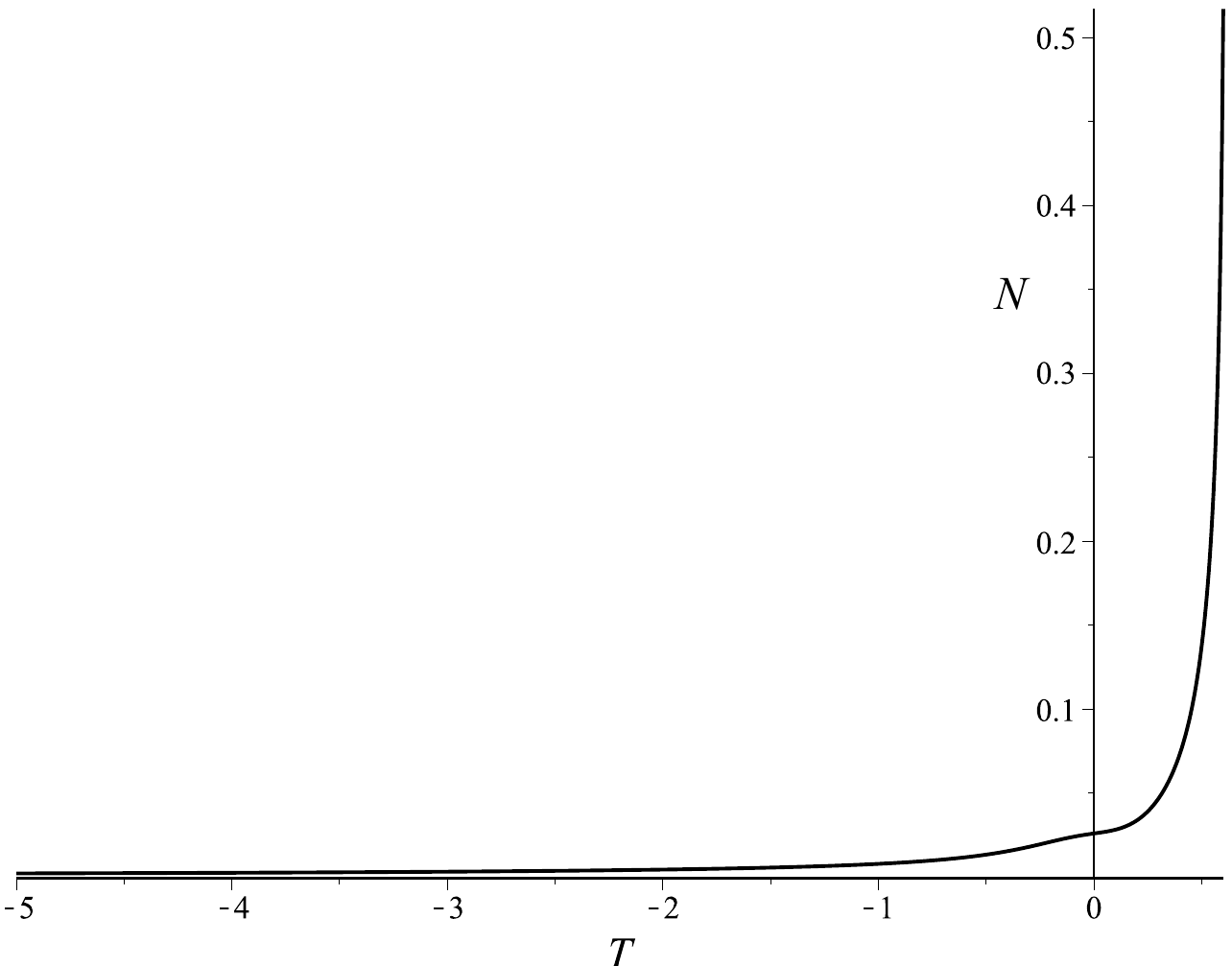}}
\end{tabular}
\end{center}
\caption{\small{Another evolution of the two-sphere scale factor ($\pc$), the Yang-Mills potential ($\potb$), and the Yang-Mills electric field ($\elecb$) for both classical and holonomy corrected evolution equations. In the quantum corrected case the volume of two-spheres does not shrink to zero, as may be seen in fig. (a), indicating the quantum bounce. Gravitational holonomy corrections and Yang-Mills Wilson loop corrections.}}
\label{fig:2_tabular_YM}
\end{framed}
\end{figure}
By comparing figures \ref{fig:1_tabular_YM} and \ref{fig:2_tabular_YM} to figures \ref{fig:1_tabular} and \ref{fig:2_tabular} we note that implementing Wilson loop corrections makes very little difference on the system. A careful analysis shows that the Wilson loop corrections lowers the $\pc$ curve slightly, hence impedes the quantum bounce, but the effect is negligible.

\section{Concluding remarks}
In this manuscript we studied homogeneous Einstein Yang-Mills black hole interiors (of topology $R\times S^{2}$) subject to low-order holonomy corrections inspired by loop quantum gravity. Specifically the evolution was performed in the ``improved quantization'' scheme known as the $\bar{\mu}^{\prime}$ scheme \cite{ref:mubarprime}. The classical solutions' two-sphere volume tend to zero as the evolution progresses, whereas the holonomy corrected evolution experiences a ``quantum bounce'' as is also found in holonomy corrected vacuum black holes. Past the classical singular point, for long-time corrected evolution, the volume of two-spheres experiences damped oscillations, asymptoting to a constant. There is a similarity in this long-time behavior to holonomy corrected pure vacuum black holes of spherical as well as non-trivial topologies. It is possible that this long term behavior is therefore universal. The $R$ sector on the other hand expands indefinitely. Interestingly, we find that the magnitude of the Yang-Mills induced electric field grows monotonically without bound on the far side of the bounce. This illustrates the complexity of the nonlinear nature of gravity and the Yang-Mills field where the electric field can be sourced by this nonlinearity. Implementing finite plaquette size corrections to the Yang-Mills field Wilson loop does not introduce a great change to the system.

\vspace{0.7cm}
\section*{Acknowledgments}
The authors are grateful to M.J. Desrochers (UBC) and J.T.G. Ghersi (currently at Perimeter Institute) for helpful discussions related to this project. We also thank H. Trottier (SFU) for clarifying points regarding lattice gauge theory. We would also like to thank the anonymous referee for constructive comments which have made the manuscript clearer.

\PRLsep
\vspace{-0.080cm}

\setcounter{equation}{0}
\renewcommand\theequation{A.\arabic{equation}}

\section*{Appendix - Holonomy corrections: brief review}
As briefly mentioned previously, the connection is an ill defined variable in the quantum theory and must be replaced by the holonomized (Wilson) loop of the connection. However, the quantum theory also predicts a finite spectrum for the area operator and hence the loop may not be shrunk down to be of vanishing size. So, one may develop a semi-classical theory by taking the classical Hamiltonian and replacing all instances of the connection $A^i_{~a}$ with its holonomized loop. Instead of shrinking the loop to zero size, we shrink it to a small finite size and this gives rise to the holonomy correction. Full details may be found in a combination of \cite{ref:schwartzbook} and \cite{ref:chiouKS}

The holonomy of the connection along some path $\alpha$ may be written as
\begin{equation}
h_\alpha = \exp \bigg[ {\int_\alpha \tau_i A^i_{~a} dx^a} \bigg]\,. \nonumber
\end{equation}
It is assumed that the connection is constant on a small enough interval, which is considered the order of the Planck length. Hence, the holonomy of distance $\delta$ in the $\partial_a$ direction is
\begin{align}
h^{(\delta)}_a &\approx \exp \bigg[ \delta\, \tau_i A^i_{\;a}    \bigg]  \label{eq:holo}\\
&=\mathbbm{1} \cos \bigg( \frac{\delta\, A^j_{a}}{2} \bigg) + 2~ \tau_j \sin\bigg( \frac{\delta\, A^j_{~a}} {2}  \bigg) \,. \nonumber
\end{align}

Now we construct the holonomy in a closed loop (see figure \ref{fig:holonomyplaquette}), $\alpha = \Box_{ab}$ such that
\begin{align}
h^{(\delta)}_{\Box_{ab}}&= h^{(\delta)}_a  h^{(\delta)}_b h^{(\delta) -1}_a h^{(\delta)-1}_b  \nonumber \\
&= \exp \Big[\delta^2 (\partial_{a} A_b - \partial_b A_a + [A_a, A_b] )  +\mathcal{O}(\delta^3) \Big]\nonumber\\
&= \exp \Big[\delta^2  F^k_{ab} \tau_k  +\mathcal{O}(\delta^3) \Big]\nonumber\\
&= \mathbbm{1}  +\tau_k  \delta^2 F^k_{ab}  +\mathcal{O}(\delta^3) ~~.
\end{align}
Hence,
\begin{equation}
F^k_{\;ab} \approx  -2 \frac{\mathrm{tr} \left[\tau_{k}(h_{\Box_{ab}} - \mathbbm{1})\right]}{\delta^2}~~.
\end{equation}

\begin{figure}[!ht]
\begin{center}
\includegraphics[width=0.50\textwidth, height=0.48\textwidth, clip,viewport=80 200 520 590]{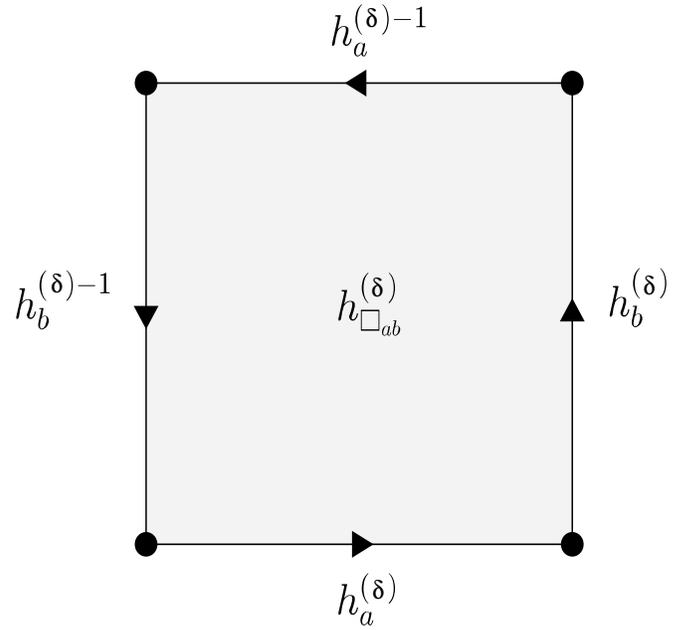}
\caption{\small{Illustration of the holonomized loop $h^{(\delta)}_{\Box_{ab}}$ which brings  the connection $A^i_{~a}$ in a small loop of side length $\delta$.}}\label{fig:holonomyplaquette}
\end{center}
\end{figure}

Now, the above trace can be computed using Eq. (\ref{eq:holo}) and it evaluates to
\begin{equation}
\mathrm{tr}[\tau_k (h_{\Box_{ab}} - \mathbbm{1}_{ab} ) ] = -\epsilon_{k i j} \frac{\sin (\delta\, A^i_{\;a})\sin (\delta\, A^j_{\;b})}{2}
\end{equation}
and so
\begin{equation}
F^k_{\;ab} = \epsilon^{k}_{\;ij} \frac{\sin(\delta\, A^i_{\;a})\sin(\delta\, A^j_{\;b})}{\delta^{2}}~.
\end{equation}
whereas classically, under the small loop approximation, we had that
\begin{equation*}
F^k_{\;ab} = \epsilon^{k}_{\;ij} A^i_{\;a} A^j_{\;b},
\end{equation*}
so the holonomy correction can be summarized as the replacement
\begin{equation}
A^i_{\; a} \rightarrow \frac{\sin (\delta\, A^i_{\;a} )}{\delta}~~,
\end{equation}
such that in the limit where $\delta \rightarrow 0$, we regain the classical case.


\PRLsep
\vspace{-0.80cm}
\linespread{0.6}
\bibliographystyle{unsrt}

\begin{thebibliography}{10}
\setlength{\itemsep}{0mm}
{\small{
\bibitem{ref:astrobh1}
S.-N. Zhang,
in \newblock{\em Astronomy Revolution: 400 Years of Exploring the Cosmos}, (D.G. York, O. Gingerich, and S.N. Zhang  eds.) Taylor \& Francis Group LLC/CRC Press (2011).

\bibitem{ref:astrobh2}
L. Rezzolla,
in \newblock{\em Astrophysical Black Holes}, Lecture Notes in Physics {\bf 905} (F. Haardt, V. Gorini, U. Moschella, A. Treves, and M. Colpi  eds.) Springer International Publishing, Switzerland (2016).

\bibitem{ref:astrobh3}
P. Shastri,
\newblock{\em Resonance} {\bf 22}(3) 237 (2017).

\bibitem{ref:gravwave}
D. Castelvecchi,
\newblock{\em Nature} {\bf 530} 261 (2016).


\bibitem{ref:oppsnyd}
J.R. Oppenheimer and H. Snyder,
\newblock{\em Phys. Rev.} {\bf 56} 455 (1939).

\bibitem{ref:isham}
C. Isham,
{\em arXiv:gr-qc/9510063} GR-14 plenary lecture, Florence, Italy (1995).

\bibitem{ref:smolin}
L. Smolin,
\newblock{\em arXiv:hep-th/0507235}.

\bibitem{ref:diffeocausalset}
J. Henson,
\newblock{\em arXiv:gr-qc/0601121v2} (2006).

\bibitem{ref:ambjornCDT}
J. Ambjorn, A. Goerlich, J. Jurkiewicz, and R. Loll,
in \newblock{\em Springer Handbook of Spacetime},  (A. Ashtekar and V. Petkov  eds.) Springer-Verlag, Berlin (2014).

\bibitem{ref:backindLQG}
A. Ashtekar and J. Lewandowski,
\newblock{\em Class. Quant. Grav.} {\bf 21} R53 (2005).

\bibitem{ref:lehtinen}
S.-L. Lehtinen,
\newblock{\em Introduction to Loop Quantum Gravity}, Master's thesis, Imperial College London (2012).

\bibitem{ref:modestobhsing2}
L. Modesto,
\newblock {\em Proc. XVII SIGRAV} (2006).

\bibitem{ref:AandB}
A. Ashtekar and M. Bojowald,
\newblock {\em Class. Quant. Grav.} {\bf 26} 391 (2006).

\bibitem{ref:BandV}
C.~ G. Boehmer and K. Vandersloot,
\newblock {\em Phys. Rev.} {\bf D76} 104030 (2007).

\bibitem{ref:modestodust}
L. Modesto,
\newblock{\em Int. J. Theor. Phys.} {\bf 47} 357 (2008).

\bibitem{ref:chiusing}
D.-W. Chiou,
\newblock{\em Phys. Rev.} {\bf D78} 064040 (2008).

\bibitem{ref:CGPint}
M.~ Campiglia, R. Gambini, and J. Pullin,
\newblock {\em AIP Conf. Proc.}: Third Mexican Meeting on Mathematical and Experimental Physics, 52 (2008).

\bibitem{ref:mysingrev}
A. DeBenedictis,
\newblock{\em Proc. Thy. Can. IV} in {\em Can. J. Phys.} {\bf 87} 255 (2009).

\bibitem{ref:taslimitehrani}
M. Taslimitehrani,
\newblock{\em Singularity Avoidance of Schwarzschild and Reissner-Nordström Black Holes in Loop Quantum Gravity}, Master's thesis, Stockholm University (supervisor H. Heydari) (2012).

\bibitem{ref:LQCsingreview}
I. Agullo and P. Singh,
in \newblock{\em Loop quantum gravity: the first 30 years}, (A. Ashtekar and J. Pullin eds.) World Scientific, Singapore (2017).

\bibitem{ref:brahma1}
M. Bojowald and S. Brahma,
\newblock{\em arXiv:1610.08850}[gr-qc].

\bibitem{ref:brahma2}
M. Bojowald and S. Brahma,
\newblock{\em Phys. Rev.} {\bf D95} 124014 (2017).

\bibitem{ref:gambinialgebra}
R. Gambini,
{\em arXiv:hep-th/9403006} (1994).

\bibitem{ref:bojowaldalgebra}
M. Bojowald, S. Brahma, and J. D. Reyes,
\newblock{\em Phys. Rev.} {\bf D92} 045043 (2015).

\bibitem{ref:achouralgebra}
J. B. Achour, S. Brahma, and A. Marciano,
\newblock{Phys. Rev.} {\bf D96,} 026002 (2017).

\bibitem{ref:tibrewalaalgebra}
R. Tibrewala,
\newblock{\em Class. Quant. Grav.} {\bf 31} 055010 (2014).

\bibitem{ref:simplefix}
M. Campiglia, R. Gambini, J. Olmedo, and J. Pullin,
\newblock{\em Class. Quant. Grav.} {\bf 33} 18LT01 (2016).

\bibitem{ref:achouralgebra2}
J. B. Achour and S. Brahma,
\newblock{\em arXiv:1712.03677}[gr-qc].

\bibitem{ref:tomasetal}
J.T.G. Ghersi, M.J. Desrochers, M. Protter, and A. DeBenedictis,
\newblock{\em arXiv:1711.04234}[gr-qc].



\bibitem{ref:ruban1}
V.A. Ruban,
{\em  Sov. Phys. JETP} {\bf 29} 1027 (1969); Reprinted: {\em Gen. Rel. Grav.} {\bf 33} 369 (2001).

\bibitem{ref:ruban2}
V.A. Ruban,
{\em  Sov. Phys. JETP} {\bf 58} 463 (1984).

\bibitem{ref:aruliah}
A. DeBenedictis, D. Aruliah, and A. Das,
\newblock{\em Gen. Rel. Grav.} {\bf 34} 365 (2002).

\bibitem{ref:zaslatsph}
O.B. Zaslavskii
{\em Phys. Rev.} {\bf D72} 067501 (2005).

\bibitem{ref:aftergood}
J. Aftergood and A. DeBenedictis,
\newblock{\em Phys. Rev.} {\bf D90} 124006 (2014).


\bibitem{ref:kantsachs}
R. Kantowski and R.K. Sachs,
\newblock{\em J. Math. Phys.} {\bf 7} 443 (1966).

\bibitem{ref:modestobhsing}
L. Modesto,
\newblock {\em Phys. Rev.} {\bf D70} 124009 (2004).

\bibitem{ref:modestolqbh}
L. Modesto,
\newblock {\em Class. Quant. Grav.} {\bf 23} 5587 (2006).

\bibitem{ref:modestoint}
L. Modesto,
\newblock{\em Adv. High Energy Phys.} {\bf 2008} 459290 (2008).

\bibitem{ref:BKD}
J. Brannlund, S. Kloster, and A. DeBenedictis,
\newblock{\em Phys.Rev.} {\bf D79} 084023 (2009).

\bibitem{ref:tibrewalaeinstmax}
R. Tibrewala,
\newblock{\em Class. Quant. Grav.} {\bf 29} 235012 (2012).

\bibitem{ref:tehrani}
M. T. Tehrani and H. Heydari,
\newblock{\em Int. J. Theor. Phys.} {\bf 51} 3614 (2012).

\bibitem{ref:schwrevisit}
A. Corichi and P. Singh,
\newblock{\em Class. Quantum Grav.} {\bf 33} 055006 (2016).

\bibitem{ref:singh2}
S. Saini and P. Singh,
\newblock{\em arXiv:1606.04932v2}[gr-qc].



\bibitem{ref:BM}
R. Bartnik and J. McKinnon,
\newblock{\em Phys. Rev. Lett.} {\bf 61} 141 (1988).

\bibitem{ref:smoller}
J.A. Smoller, A.G. Wasserman, S.-T. Yau, and J.B. McLeod,
\newblock{\em Comm. Math. Phys.} {\bf 143} 115 (1991).

\bibitem{ref:wasserman}
A. G. Wasserman,
\newblock{\em J. Math. Phys.} {\bf 41} 6930 (2000).

\bibitem{ref:dascoff}
  A.~Das and C.V.~Coffman,
 \newblock{\em J. Math. Phys.} {\bf 8} 1720 (1967).


\bibitem{ref:volgal}
M.S. Volkov and D.V. Gal'tsov,
\newblock{\em JETP Lett.} {\bf 50}(7) 346 (1989).

\bibitem{ref:breit}
P. Breitenlohner, P. Forg\'{a}cs, and D. Maison,
\newblock{\em Comm. Math. Phys.} {\bf 163} 141 (1994).

\bibitem{ref:donets}
E.E. Donets, D.V. Gal'tsov, and M.Y. Zotov,
\newblock{\em Phys. Rev.} {\bf D56} 3459 (1997).

\bibitem{ref:volkovreview}
M.S. Volkov and D.V. Gal'tsov,
\newblock{\em Phys. Rept.} {\bf 319} 1 (1999).

\bibitem{ref:galthiggs}
D.V. Gal'tsov and E.E. Donets,
\newblock{\em Compt. Rend. Acad. Sci. Serb.} {\bf IIB 321}(11) 649 (1997).

\bibitem{ref:paturhiggs}
V. Paturyan, E. Radu, and D.H. Tchrakian,
\newblock{\em Phys. Lett.} {\bf B609} 360 (2005).

\bibitem{ref:jiahiggs}
J. Jia,
\newblock{\em Can. J. Phys.} {\bf 88}(31) 189 (2010).

\bibitem{ref:winstanley1}
E. Winstanley,
\newblock{\em Class. Quant. Grav.} {\bf 16} 1963 (1999).

\bibitem{ref:breitlambda}
P. Breitenlohner, P. Forg\'{a}cs, and D. Maison,
\newblock{\em Comm. Math. Phys.} {\bf 261} 569 (2006).

\bibitem{ref:pertaxial}
O. Brodbeck, M. Heusler, N. Straumann and M.S. Volkov,
\newblock{\em Phys. Rev. Lett.} {\bf 79} 4310 (1997).

\bibitem{ref:kleiaxial}
B. Kleihaus, J. Kunz, and F. Navarro-Lerida,
\newblock{\em Phys.Rev.} {\bf D66} 104001 (2002).

\bibitem{ref:bizondyon}
P. Bizon and O.T. Popp,
\newblock{\em Class. Quant. Grav.} {\bf 9} 193 (1992).

\bibitem{ref:improved}
M. Zotov,
\newblock{\em arXiv:gr-qc/9704080v2} (1997).

\bibitem{ref:eymdilaton}
B. Kleihaus and J. Kunz,
\newblock{\em Phys. Rev.} {\bf D57} 834 (1998).

\bibitem{ref:breitchaos}
P. Breitenlohner, G. Lavrelashvili, and D. Maison,
\newblock{\em Nucl. Phys.} {\bf B524} 427 (1998).

\bibitem{ref:choptuik}
M.W. Choptuik, E.W. Hirschmann, and R.L. Marsa,
\newblock{Phys. Rev.} {\bf D60} 124011 (1999).

\bibitem{ref:hosotanidyon}
Y. Hosotani and J. Bjoraker,
\newblock{\em Proc. 6$^{\mbox{th}}$ Wigner Symp.} Istanbul, Turkey (1999).

\bibitem{ref:balakin}
A.B. Balakin, J.P.S. Lemos, and A.E. Zayats,
\newblock{\em Phys. Rev.} {\bf D93} 084004 (2016).

\bibitem{ref:criticalmaliborsk}
M. Maliborsk and O. Rinne,
\newblock{\em arXiv:1712.0458}[gr-qc] (2017).



\bibitem{ref:samarkandrev}
R. Ibadov and J. Kunz,
\newblock{Samarkand presentation,} March/April 2007:\\
\newblock{\tiny{\url{https://www.uni-oldenburg.de/fileadmin/user_upload/physik/ag/feldtheorie/download/talks/Samarkand2.pdf}}}.

\bibitem{ref:richness}
E. Winstanley,
\newblock{presentation:}\\
\newblock{\tiny{\url{https://www.newton.ac.uk/files/seminar/20051201160017001-149501.pdf}}}.



\bibitem{ref:entrev1}
K.~A. Meissner,
\newblock {\em Class. Quant. Grav.} {\bf 21} 5245 (2004).

\bibitem{ref:entrev2}
A. Ghosh and P. Mitra,
\newblock {\em Phys. Let.} {\bf B616} 114 (2005).

\bibitem{ref:immirzi}
K.~A. Meissner,
\newblock {\em Class. Quant. Grav.} {\bf 21} 5245 (2004).

\bibitem{ref:entrev3}
A. Corichi, J. D\'{i}az-Polo, and E. Fern\'{a}ndez-Borja,
\newblock {\em Class. Quant. Grav.} {\bf 24} 243 (2007).

\bibitem{ref:jacobson}
T. Jacobson,
\newblock {\em 	Class. Quant. Grav.} {\bf 24} 4875 (2007).

\bibitem{ref:highergenusent}
S. Kloster, J. Brannlund, and A. DeBenedictis,
\newblock{\em Class. Quant. Grav.} {\bf 25} 065008 (2008).

\bibitem{ref:mubarprime}
D-W. Chiou,
\newblock {\em Phys. Rev.} {\bf D76} 124037 (2007).

\bibitem{ref:musemiclass}
A. Ashtekar, T. Pawlowski, and P. Singh,
\newblock{\em Phys. Rev.} {\bf D74} 084003 (2006)

\bibitem{ref:chiouKS}
D-W. Chiou,
\newblock {\em Phys. Rev.} {\bf D78} 044019 (2008).

\bibitem{ref:schwartzbook}
M.D. Schwartz,
\newblock{\em Quantum Field Theory and the Standard Model},  Cambridge University Press, Cambridge (2014).

\bibitem{ref:miyachi}
Y. Miyachi, M. Ikeda, and T. Maekawa,
\newblock{\em Prog. Theor. Phys.} {\bf 68} 261 (1982).

\bibitem{ref:DJS}
N. Dadhich, A. Joe, and P. Singh,
\newblock{\em Class. Quant. Grav.} {\bf 32} 185006 (2015).


 }}
\end{thebibliography}

} 
\end{document}